\documentclass[sigplan,screen]{acmart}
\renewcommand\footnotetextcopyrightpermission[1]{} 

\usepackage{pgfplots}
\usepackage{pgfplotstable}
\usepackage{booktabs}
\usepackage{subcaption} 
\usepackage{algorithm}
\usepackage{listings}
\usepackage{color}

\definecolor{dkgreen}{rgb}{0,0.6,0}
\definecolor{gray}{rgb}{0.5,0.5,0.5}
\definecolor{mauve}{rgb}{0.58,0,0.82}

\usepackage[noend]{algorithmic}
\usepackage{fancyvrb}
\usetikzlibrary{external}
\pgfplotsset{compat=1.9}
\AtBeginDocument{%
  \providecommand\BibTeX{{%
    \normalfont B\kern-0.5em{\scshape i\kern-0.25em b}\kern-0.8em\TeX}}}

\setcopyright{acmcopyright}
\copyrightyear{2018}
\acmYear{2018}
\acmDOI{10.1145/1122445.1122456}

\begin{document}


\title{Elastic Deep Learning in Multi-Tenant GPU Clusters}

\author{Yidi Wu}
\email{ydwu@cse.cuhk.edu.hk}
\affiliation{%
  \institution{the Chinese University of Hong Kong}
  \city{}
  \state{}
}
\author{Kaihao Ma}
\email{khma@cse.cuhk.edu.hk}
\affiliation{%
	\institution{the Chinese University of Hong Kong}
	\city{}
	\state{}
}
\author{Xiao Yan}
\email{xyan@cse.cuhk.edu.hk}
\affiliation{%
	\institution{the Chinese University of Hong Kong}
	\city{}
	\state{}
}
\author{Zhi Liu}
\email{zliu@cse.cuhk.edu.hk}
\affiliation{%
	\institution{the Chinese University of Hong Kong}
	\city{}
	\state{}
}
\author{Zhenkun Cai}
\email{zkcai@cse.cuhk.edu.hk}
\affiliation{%
	\institution{the Chinese University of Hong Kong}
	\city{}
	\state{}
}
\author{Yuzhen Huang}
\email{yzhuang@cse.cuhk.edu.hk}
\affiliation{%
	\institution{the Chinese University of Hong Kong}
	\city{}
	\state{}
}
\author{James Cheng}
\email{jcheng@cse.cuhk.edu.hk}
\affiliation{%
	\institution{the Chinese University of Hong Kong}
	\city{}
	\state{}
}

\author{Han Yuan}
\email{yuanhan3@huawei.com}
\affiliation{%
	\institution{Huawei Technologies}
	\city{}
	\state{}
}
\author{Fan Yu}
\email{fan.yu@huawei.com}
\affiliation{%
	\institution{Huawei Technologies}
	\city{}
	\state{}
}

\pagestyle{plain} 
\begin{abstract}

We study how to support elasticity, i.e., the ability to dynamically adjust the parallelism (number of GPUs), for deep neural network (DNN) training. Elasticity can benefit multi-tenant GPU cluster management in many ways, e.g., achieving various scheduling objectives (e.g., job throughput, job completion time, GPU efficiency) according to cluster load variations, maximizing the use of transient idle resources, performance profiling, job migration, and straggler mitigation. However, existing parallelism adjustment strategies incur high overheads, which hinder many applications from making effective use of elasticity. We propose EDL to enable low-overhead elastic deep learning with a simple API. We present techniques that are necessary to reduce the overhead of parallelism adjustments, such as stop-free scaling and dynamic data pipeline. We also demonstrate that EDL can indeed bring significant benefits to the above-listed applications in GPU cluster management.

\end{abstract}


\maketitle

\section{Introduction}\label{sec:intro}

Due to the huge success of deep learning (DL), many organizations have built large GPU clusters for deep neural network (DNN) training. A GPU cluster typically serves many concurrent users. Users submit training jobs and the resource requirements (e.g., the number of GPUs) to the cluster. These multi-tenant GPU clusters are usually managed by traditional cluster managers (e.g., YARN~\cite{VavilapalliMDAKEGLSSSCORRB13cloud}, Mesos~\cite{HindmanKZGJKSS10nsdi}) or schedulers tailored for GPU clusters (e.g., Optimus~\cite{PengBCWG18eurosys}, Gandiva~\cite{XiaoBRSKHPPZZYZ18osdi}, Tiresias~\cite{GuCSZJQLG19nsdi}), with scheduling objectives such as \textit{high throughput, high GPU efficiency~\footnote{Here, \textit{throughput} is the average number of training samples processed per second. Let $t(p)$ be the average per-GPU throughput of a job using $p$ GPUs, and $p^*={\it argmax}_{p} t(p)$. \textit{GPU efficiency} is defined as $t(p)/t(p^*)$, which is an indicator of how close the current average per-GPU throughput (using $p$ GPUs) is to the optimal one (using $p^*$ GPUs).}, short job completion time (JCT), and good responsiveness for small jobs~\footnote{Following Tiresias~\cite{GuCSZJQLG19nsdi}, we define \textit{job size} as \textit{parallelism$\times$running time}.}}.

Through an analysis of the trace data from Microsoft's production GPU cluster~\cite{JeonVPQXY19atc}, we found that~\textit{elasticity, the ability to adjust the parallelism (i.e., the number of GPUs) of a DNN training job}, is beneficial to multi-tenant GPU cluster management in many aspects. \textit{First, elasticity can adjust the trade-off between throughput and GPU efficiency~\footnote{For the same job, using more GPUs often increases throughput but decreases GPU efficiency.}, enabling us to more flexibly adapt to cluster load variations.}  GPU clusters are fully loaded sometimes but  under-utilized at other times. With elasticity, DNN training jobs can be scaled out  (i.e., increase the parallelism) to achieve high throughput when the cluster is not busy, and scaled in to improve GPU efficiency when the cluster is heavily loaded. \textit{Second, elasticity allows transient idle resources to be effectively utilized}.  Transient idle resource are common in GPU clusters (even in peak hours as revealed in Microsoft's trace data) and most of the idle intervals are short (lasting for a few minutes). Jobs can be scaled out to make use of these transient idle resources, and scaled in to return the resources when they need to be re-allocated to other jobs. \textit{Third, elasticity can help enforce priority and improve the responsiveness for small jobs}. It is reported that small jobs are usually used for program correctness checking, parameter configuration tuning, and network architecture search, for which quick response is critical~\cite{GuCSZJQLG19nsdi, XiaoBRSKHPPZZYZ18osdi}. We found from the trace data that there are many small jobs as well as jobs of various sizes queuing during peak hours. Thus, elasticity can be used to scale in large or low-priority jobs in order to allocate sufficient resources to small or high-priority jobs, which helps prevent head-of-line blocking and improve responsiveness and average JCT. Finally, important scheduling operations such as \textit{straggler mitigation, performance profiling, and job migration} can be easily implemented based on elasticity~(\S\ref{sec:usecase}).

The main challenge of supporting elasticity is to make its overhead sufficiently low. Parallelism can be trivially adjusted by \textit{stop-resume}~\cite{PengBCWG18eurosys}, which checkpoints a job and restarts it with the desired parallelism, but the job typically needs to be stopped for more than 30 seconds. This overhead puts a limit on the frequency of parallelism adjustments, which in turn limits the application scenarios of elasticity. For example, Optimus~\cite{PengBCWG18eurosys} uses \textit{stop-resume} only every 10 minutes in order to amortize the large overhead. In the case of using transient idle GPUs, the overhead of stop-resume is also too high for us to gain a performance improvement~(\S\ref{subsec:benifits of elasticity}). In addition, the system should make the mechanism of elasticity transparent to users and the changes that need to be made to users' programs should be minimal.

We propose \textbf{EDL} to support \textit{elastic deep learning} in multi-tenant GPU clusters. EDL is a light-weight coordination layer between a cluster scheduler and a DL framework. Similar to Horovod~\cite{SergeevBcorr18}, EDL delegates single-machine execution to the underlying DL framework (e.g., TensorFlow~\cite{AbadiBCCDDDGIIK16osdi}, PyTorch~\cite{Paszke2017automatic}, MXNet~\cite{ChenLLLWWXXZZ15corr}). The DL framework only needs to retrieve the meta data of a block of training data from EDL and notifies EDL after finishing a mini-batch. EDL can be used as a simple plug-in to different DL frameworks and maintains good usability, i.e., users only need to add a few lines to their script to enjoy elasticity and EDL hides all the details (e.g., dynamic parallelism adjustments) from users. The scheduler can instruct EDL to remove/add any worker for a training job using a simple API, e.g., \textit{sclae\_in()} and~\textit{sclae\_out()}. Most critically, EDL significantly reduces the overhead of elasticity compared with stop-resume.

EDL is designed to ensure both correctness and efficiency when any worker may join/leave a job at any time. In EDL, each job is managed by a leader process and EDL uses a distributed transaction-based mechanism for fast leader election. To scale out, EDL proposes \textit{stop-free scaling}, which allows existing workers to continue training while newly added workers are being prepared for execution. This hides most of the scaling overhead. To scale in, EDL uses \textit{graceful exit} to remove workers at the end of a mini-batch training with negligible overhead. For data preparation, EDL uses a \textit{dynamic data pipeline} to assign blocks of data to workers in an on-demand fashion and leverages data pre-fetching to avoid starvation of GPUs. The data pipeline also ensures that the training goes over a dataset once without repetition and omission in each epoch.

We conducted extensive experiments to validate the performance of EDL. In trainings that use static parallelism without scaling, EDL achieves similar throughput and good scalability as Horovod. Compared with stop-resume, EDL reduces the overhead of scaling out by an order of magnitude and has negligible overhead for scaling in. In addition, we showed that using EDL brings significant benefits to applications such as straggler mitigation, performance profiling, and job migration. With some simple modifications, we enabled Tiresias~\cite{GuCSZJQLG19nsdi}, a state-of-the-art DL scheduler, to efficiently apply elasticity in DNN job scheduling, which achieves a reduction in the average JCT by 89.5\%. 


In~\S\ref{sec:background and motivation}, we give the motivation of the work. In~\S\ref{sec:api}~and~\ref{sec:design}, we present the API and the system design. In~\S\ref{sec:usecase}, we discuss the use cases of EDL. In~\S\ref{sec:experiment}, we report the experimental results. In~\S\ref{sec:related}~and~\ref{sec:end}, we give the related work and conclusions.

\section{Motivation}\label{sec:background and motivation}

We first give some background on distributed DNN training. Then, we motivate our work by presenting the benefits of elasticity and highlighting the challenges of the work.

\subsection{Background}\label{subsec:background}

A DNN model is trained by going over a dataset many times (called epochs), and in each epoch the dataset is randomly shuffled and partitioned into a number of mini-batches. For each mini-batch, the model is updated using stochastic gradient descent (SGD), or its variants such as Adam and AdaGrad, with $w^{(t+1)}=w^{(t)}-\frac{\eta_t}{|\mathcal{B}_t|}\sum_{i\in\mathcal{B}_t} \nabla f(x_i, w^{(t)})$, where $w^{(t)}$ is the current model and $\mathcal{B}_t$ contains the training samples of the mini-batch. As calculating the gradient $\nabla f(x_i, w^{(t)})$ involves computation-intensive kernels such as matrix multiplication, DNN training is usually conducted on GPUs.  

Due to the growing volume of data and the high complexity of DNN models (e.g., ResNet~\cite{HeZRS16cvpr}, VGG~\cite{SimonyanZ15iclr}, Inception~\cite{SzegedyVISW16cvpr}), DNN training usually cannot be finished within a reasonable time on a single GPU and thus distributed training on multiple GPUs offers a good alternative. Among the various distributed training schemes, \textit{synchronous data-parallel} is the most popular one\cite{KimYPCJHLJC19eurosys}, which partitions a dataset among the GPUs and each GPU (i.e., a worker) calculates the gradient for some training samples in parallel. When all workers finish the gradient computation in a mini-batch, the local gradient from the workers are aggregated and then added to the model before the next mini-batch starts.

\textit{Allreduce}~\cite{PatarasukY09jpdc, NCCL} is a popular protocol for coordinating model updates from distributed workers and has been widely adopted in TensorFlow, PyTorch, MXNet and Horovod due to its simplicity and high network efficiency. We present the implementation of Ring-Allreduce as follows. Workers form a ring communication topology and each worker communicates only with its two neighbors on the ring. When one gradient tensor is ready, each worker sends, receives and aggregates $1/N$ (where $N$ is the number of workers) of the tensor to the adjacent worker in a round-robin fashion in each step. After $N-1$ steps, each worker has $1/N$ of the tensor that aggregates the updates from all workers. In the next $N-1$ step, each worker passes its aggregated part of the parameters along the ring such that the gradient on all workers will be updated.  \textit{Parameter server}~\cite{HuangJWCYYLGC18pvldb, LiAPSAJLSS14osdi, ZhangZXDHLHWXX17atc, XingHDKWLZXKY15kdd} is also widely used in distributed machine learning, which provides a key-value interface for model update/lookup. However, configuration is more complicated for parameter server as performance strongly depends on the number and location of the servers as well as the skewness of tensor sizes in the models\cite{GuCSZJQLG19nsdi, PengBCWG18eurosys, PengZCBYLWG19sosp, JayarajanWGFP2019sysml}.

Horovod~\cite{SergeevBcorr18} is the state-of-the-art framework for distributed DNN training based on Allreduce. It delegates single machine execution to existing deep learning frameworks and adopts the synchronous data-parallel computation model. In a Horovod job, a leader process coordinates the order and granularity of gradient synchronization among workers.


\subsection{The Benefits of Elasticity}\label{subsec:benifits of elasticity}

We show the benefits of elasticity for multi-tenant GPU cluster management from observations in our experiments and the trace data from Microsoft~\cite{Philly-traces}. The trace data contains scheduling events (e.g., job submission/finish time) and brief descriptions of jobs (e.g. user id, number and location of allocated GPUs) collected over two months from Microsoft's production GPU cluster (with approximately 2,300 GPUs).

\vspace{-1mm}

\begin{figure}[!t]	
	\centering 
	\begin{minipage}[b]{0.23\textwidth}
		\centering
		\includegraphics[width=\textwidth]{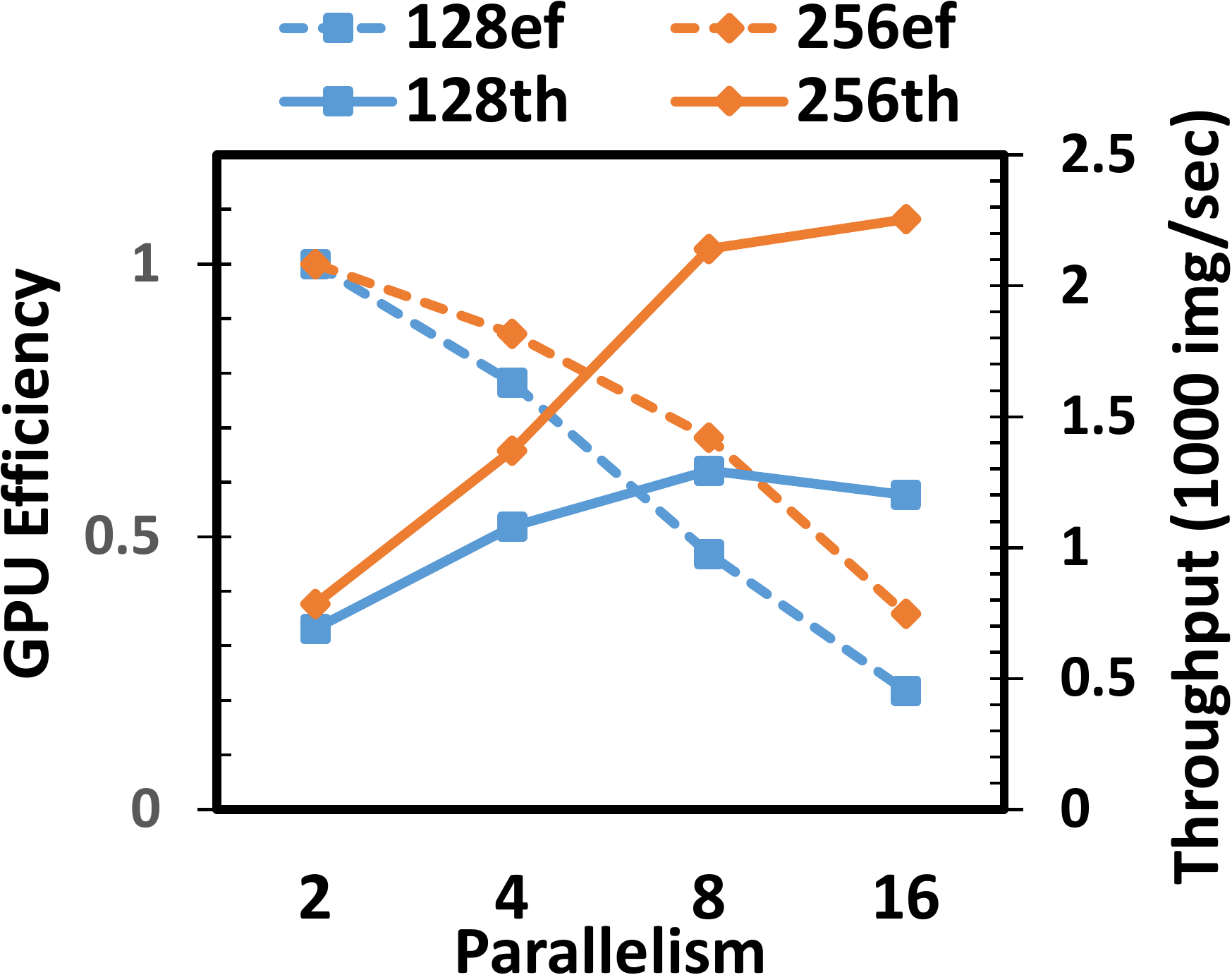}
		\subcaption{ResNet50}
	\end{minipage}
	\begin{minipage}[b]{0.23\textwidth}
		\centering
		\includegraphics[width=\textwidth]{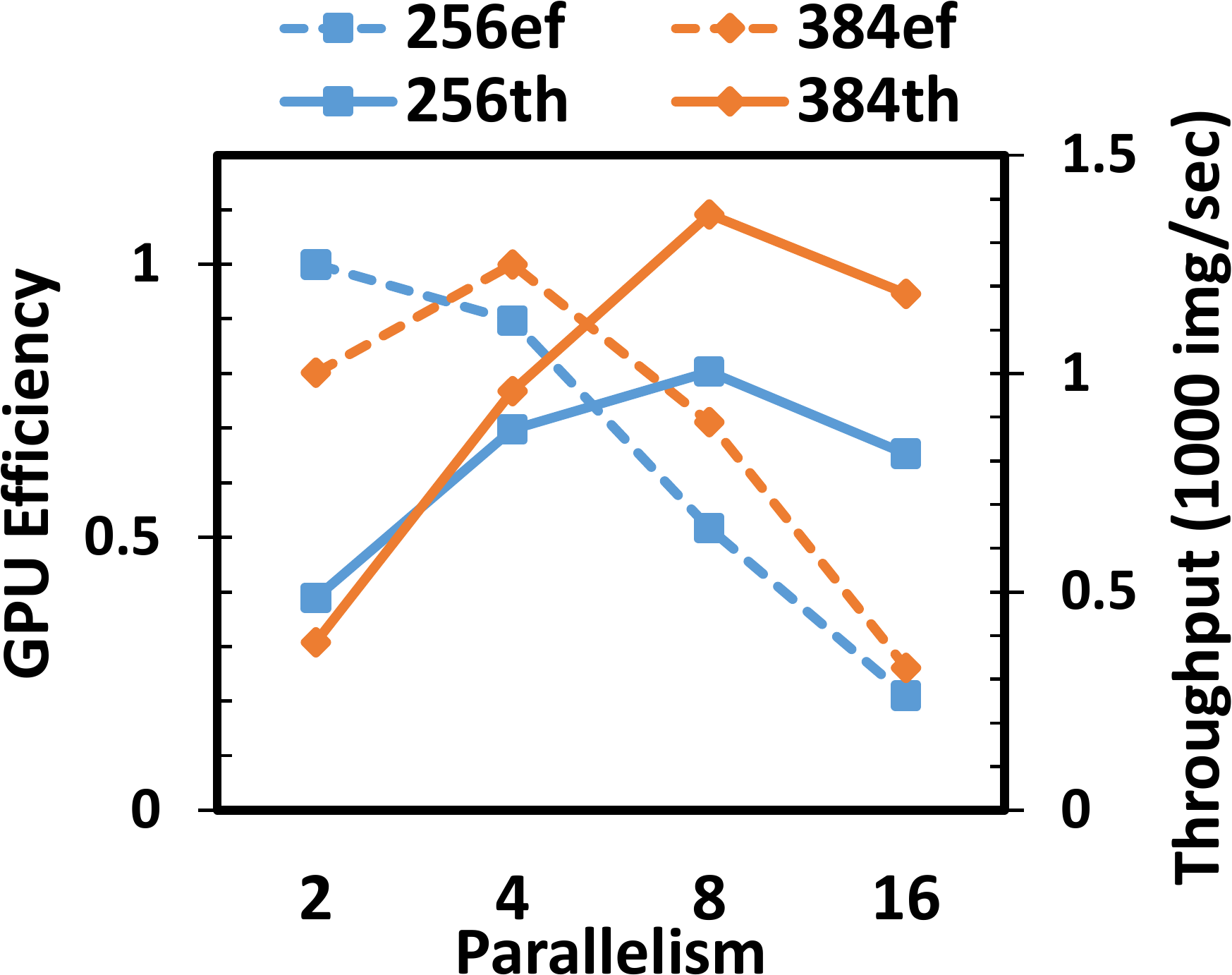}
		\subcaption{VGG16}
	\end{minipage}
	\vspace{-3mm}
	\caption{Throughput and GPU efficiency vs. parallelism}
	\label{fig:throughput efficiency parallelism}
	\vspace{-1mm}
\end{figure}

\paragraph{Adjusting the trade-off between throughput and efficiency.} Figure~\ref{fig:throughput efficiency parallelism} shows the throughput (th) and GPU efficiency (ef) of training VGG19 and ResNet50 using different batch sizes \footnote{The batch size refers to the aggregate batch size of all GPUs running a job.} ($b$) and parallelism. The throughput of ResNet50 increases with the parallelism but the gain diminishes, while the GPU efficiency decreases with the parallelism. This is because distributed training needs to pay a higher communication cost under a larger parallelism. The throughput of VGG19 even drops when the parallelism exceeds 8 GPUs because of the high communication cost caused by the large model. Under a batch size of 384, the best GPU efficiency of training VGG19 is achieved with 4 GPUs because the intermediate result (activations of the layers) takes too much memory on each GPU under smaller parallelism, which degrades the computing throughput of the kernels due to insufficient cache space.


The results show that DNN training jobs can usually run with a range of parallelism and the trade-off between throughput and GPU efficiency can change significantly with parallelism. Thus, elasticity can be used to dynamically adjust the parallelism of DNN training jobs according to different objectives of cluster scheduling.

\begin{figure}[!t]	
	\centering 
	\begin{minipage}[b]{0.225\textwidth}
		\centering
		\includegraphics[width=\textwidth]{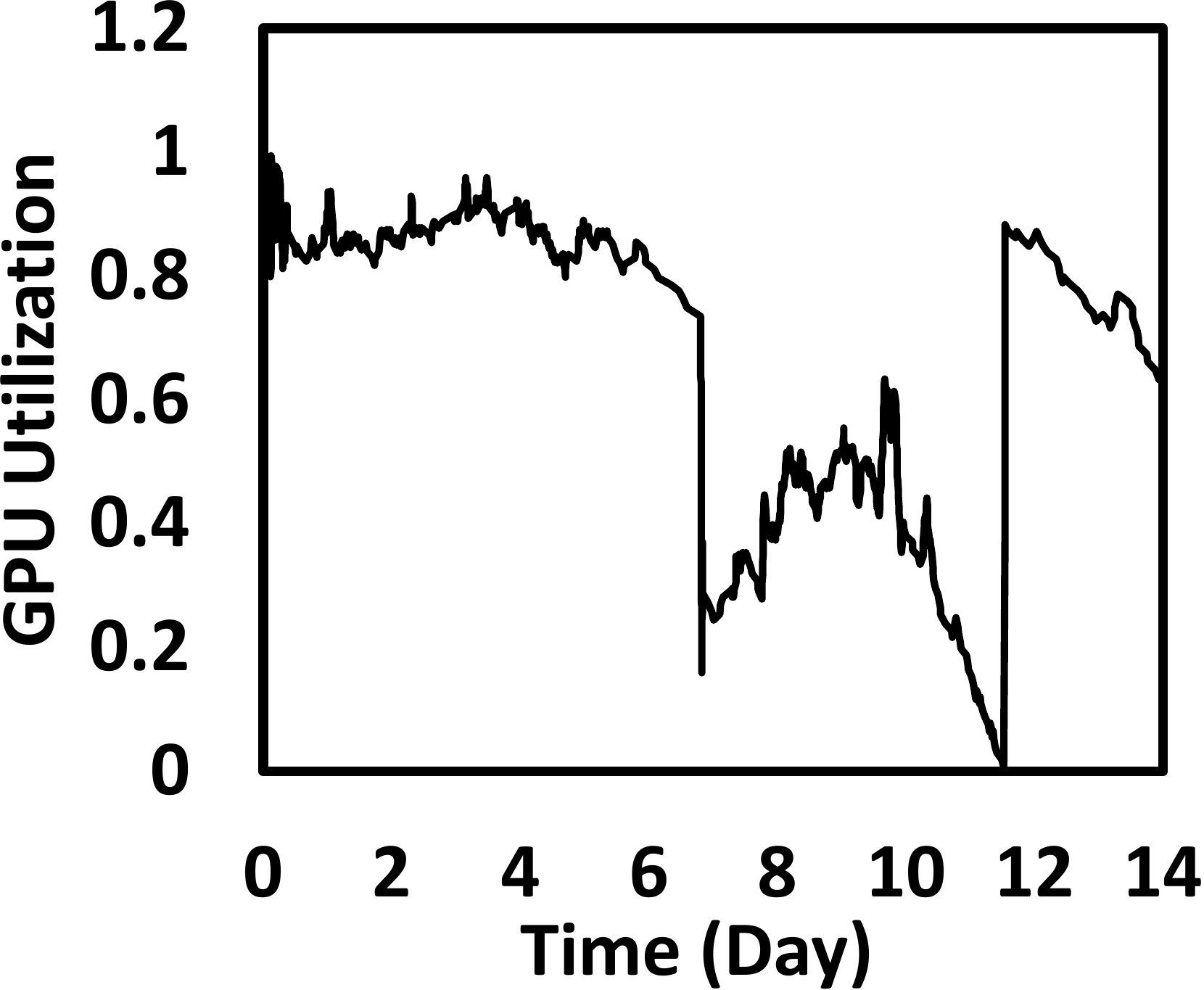}
		\subcaption{Cluster load over time}\label{fig:load}
	\end{minipage}
	\begin{minipage}[b]{0.213\textwidth}
		\centering
		\includegraphics[width=\textwidth]{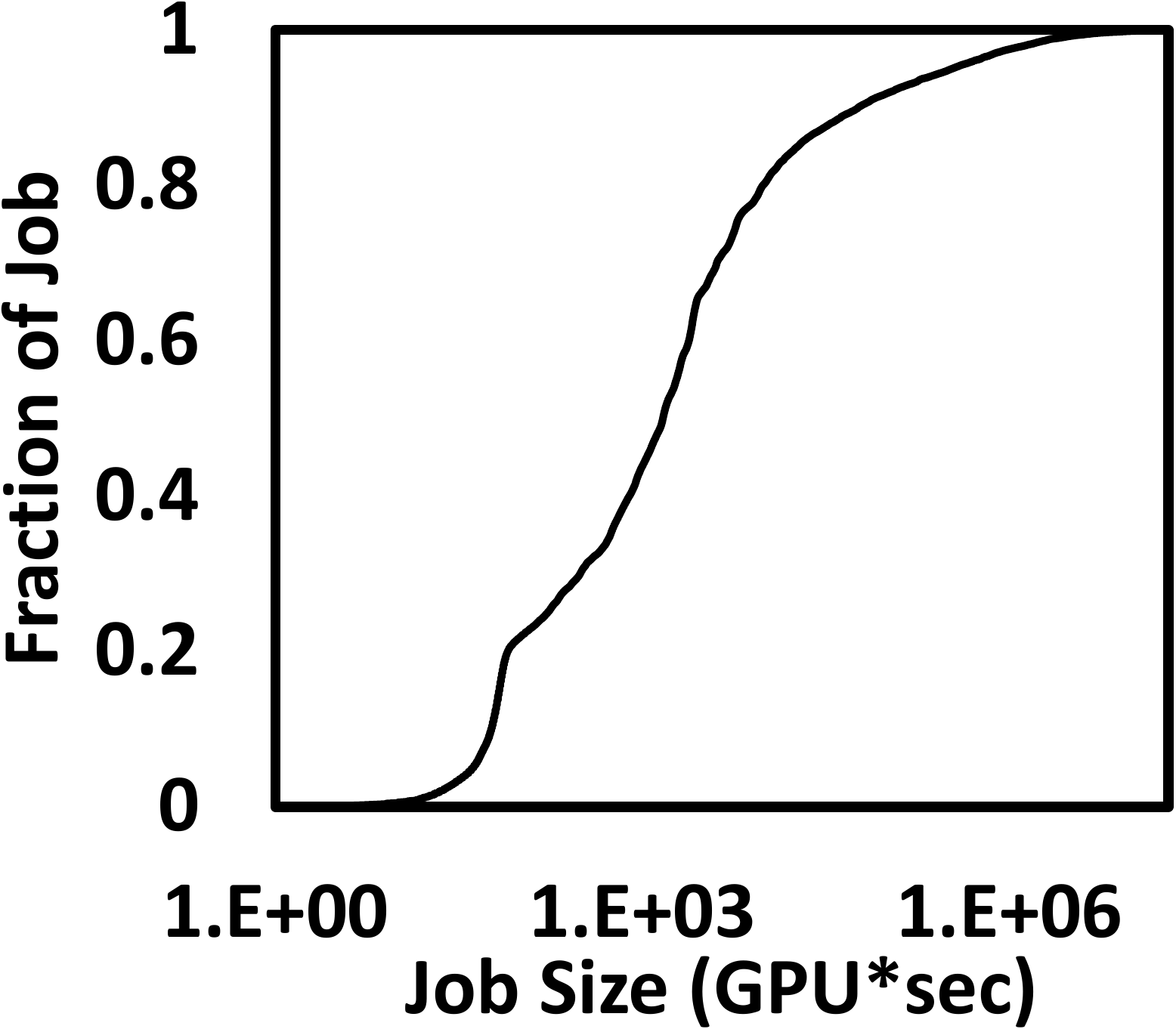}
		\subcaption{Job size CDF}\label{fig:size}
	\end{minipage}
	\vspace{-3mm}
	\caption{Variations of cluster load and job sizes}
	\label{fig:load and size}
\end{figure}

\paragraph{Improving cluster utilization and JCT.} We plot the changes in the load of the Microsoft cluster over time in Figure~\ref{fig:load}. The cluster is almost fully loaded in some periods and many jobs are queuing to be processed, while in other periods the cluster load is relatively low. 
We also plot the cumulative distribution function (CDF) of the sizes of all jobs over the two-month period in Figure~\ref{fig:size}, which shows that there exists a large variation in job sizes. Among the jobs, the 20th percentile takes 85 GPU*sec while the 90th percentile takes 58,330 GPU*sec. 

Based on the above findings, elasticity is useful in improving cluster utilization and JCT in the following ways: (1)~when the cluster load is high, scaling in large jobs (or low-priority jobs) as to improve GPU efficiency, while the GPUs freed from the scale-in can be used to run small jobs (or high-priority jobs) that are queuing as to reduce their JCT; and (2)~when the cluster load is low, scaling out jobs (e.g., high-priority ones) to make fuller utilization of the cluster and improve throughput and JCT. Our experiments in \S\ref{sec:experiment} show that by enabling elasticity, the average JCT is reduced by 89.5\%.

\begin{figure}[t]
	\centering
	\includegraphics[width=0.225\textwidth]{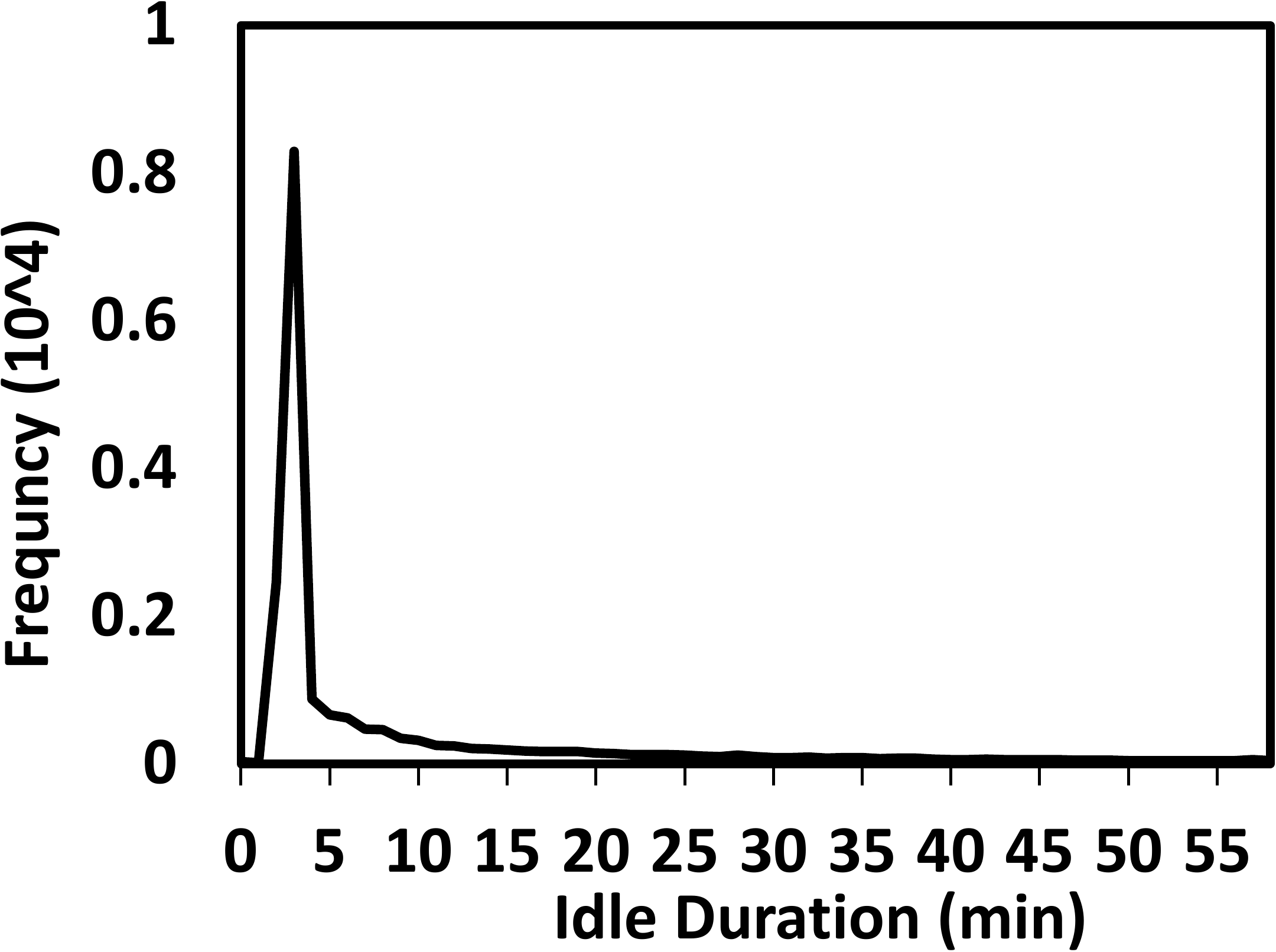}
	\vspace{-2.5mm}	
	\caption{Distribution of idle GPU intervals}
	\label{fig:idle interval}
	\vspace{-1mm}	
\end{figure}

\vspace{-1mm}	
\paragraph{Utilizing transient idle resources.} We define the \textit{idle interval} of a GPU as the time elapsed between the finish of the previous job and the start of the next job on the GPU. We plot the frequency distribution of the idle intervals of the GPUs from the trace data in Figure~\ref{fig:idle interval}. The idle intervals follow a power law distribution and the majority are short intervals. Specifically, 39.62\% of the idle intervals are less than 4 minutes, which takes up 41.5\% of the idle resources during peak hours (when $>$90\% of the GPUs are occupied).

Elasticity helps utilize transient idle resources by scaling out a job when some GPUs become idle and scaling in the job when other jobs need to use these GPUs later. However, stop-resume based elasticity cannot effectively utilize transient idle resources due to its high overhead. We used stop-resume to adjust the parallelism of a job from 1 GPU to different number of GPUs. The overhead ranges from 40 to over 80 seconds \footnote{The scaling overhead increases with parallelism as TensorFlow initializes the GPU devices in one machine sequentially.} as shown Figure~\ref{fig:scaling}. This renders stop-resume based elasticity impractical, which we explain using an example. Consider a job running on 4 GPUs and we have a GPU that will be idle for 4 minutes. Stop-resume needs to first adjust the parallelism from 4 to 5 and then back to 4. Assume that each parallelism adjustment takes 30 seconds, training is conducted with 5 GPUs for at most 3 minutes. Thus, the effective training time is at most (5 GPUs * 180 sec) $=$ 900 GPU*sec. In contrast, the effective training time is (4 GPUs * 240 sec) $=$ 960 GPU*s if we do not use the idle GPU at all. In fact, the high overhead of stop-resume means that scaling can only be conducted infrequently, which limits its ability to adapt to dynamic resource availability and job requirements (e.g., resource demands, priority).  

\section{System Architecture and APIs}\label{sec:api}

\begin{figure}[!t]	
	\centering 
	\includegraphics[width=0.28\textwidth]{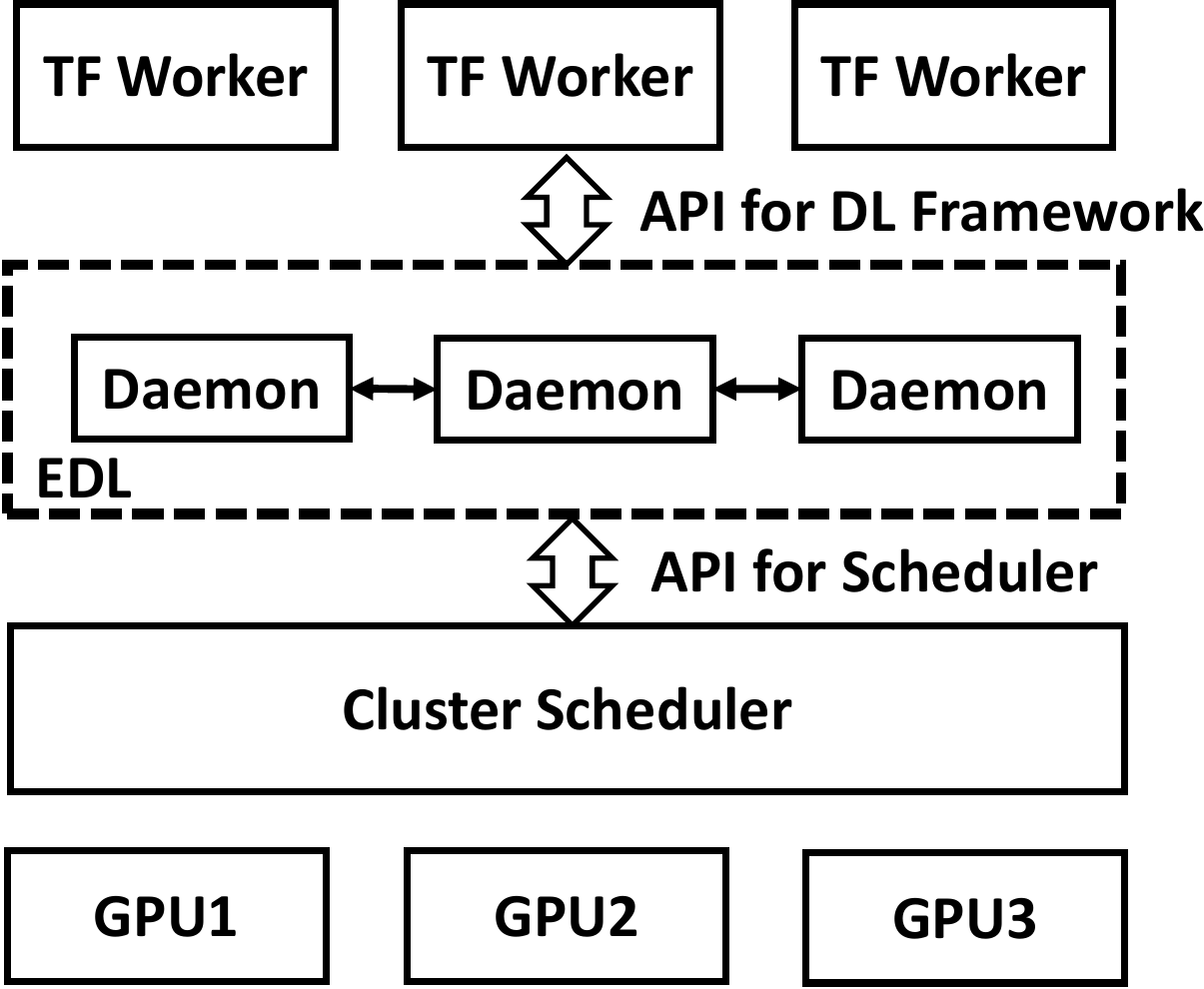}
	\caption{The positioning of EDL}
	\label{fig:system architecture}
\end{figure}

We focus on \textit{data-parallel, synchronous training} as it is the dominant paradigm of distributed DL~\cite{AbadiBCCDDDGIIK16osdi, ChenLLLWWXXZZ15corr, Paszke2017automatic, SergeevBcorr18, KimYPCJHLJC19eurosys}. As users of DL are mostly familiar with popular frameworks such as TensorFlow, PyTorch and MXNet, it would be desirable if the core logic of elasticity can be shared among different DL frameworks. The shared component could be a new elastic communication library like Nvidia NCCL~\cite{NCCL} or parameter server~\cite{XingHDKWLZXKY15kdd, HuangJWCYYLGC18pvldb, LiAPSAJLSS14osdi, ZhangZXDHLHWXX17atc, XingHDKWLZXKY15kdd}. However, supporting elasticity not only requires synchronizing the model among an elastic set of processes but also involves dynamically partitioning the training data and modifying parameters such as per-GPU batch size. Thus, we design EDL as a coordination layer sitting between DL frameworks and the GPU cluster manager as shown in Figure~\ref{fig:system architecture}. The key APIs of EDL are summarized in Table~\ref{tab:API}. The cluster manager can use EDL's scheduler API to adjust the parallelism of jobs without knowing the details of the parallelism adjustment procedures. Users write training scripts using existing DL frameworks and only need to add a few lines to use EDL. This design incurs minimal change to existing infrastructures and results in good usability.

In EDL, each job is executed by a group of worker processes and each process is associated with an EDL daemon. A \textit{leader} is elected among the workers to schedule the order and granularity of gradient synchronization (the synchronization process is similar to Horovod~\cite{SergeevBcorr18}) and coordinate the parallelism adjustment (\S\ref{design:flexibility}). Each worker process is attached with one GPU and runs in the single-machine mode using a DL framework to compute gradient on some training samples for a mini-batch.  We ingest communication operators (e.g., using the Grapler graph edit APIs in TensorFlow or hooks in PyTorch) in-between the computation and accumulation of gradients. Within the communication operator, an EDL daemon sends tensor synchronization requests asynchronously to the leader. After receiving ready-to-reduce message from the leader, the EDL daemon delegates the synchronization task to a dedicated thread to avoid blocking message handling and the thread conducts gradient synchronization using communication libraries such as NCCL.

\begin{table}[t]
	\centering
	\caption{Key APIs of EDL}
	\label{tab:API}
	\scalebox{0.8}{
		\begin{tabular}{c|c}
			\toprule
			\textbf{API for cluster scheduler} &\textbf{Description} \\
			\hline
			\textit{sclae\_in}(\textit{job\_handle, rmv\_GPU\_info}) &	remove GPUs from a job\\
			\textit{sclae\_out}(\textit{job\_handle, add\_GPU\_info})	& add GPUs to a job\\
			\textit{profile}(\textit{job\_handle, min\_p, max\_p}) & profile a job\\
			\bottomrule
			\textbf{API for DL frameworks} &\textbf{Description} \\
			\hline
			\textit{elastic\_shard\_generator}() & generate the next shard's info \\
			\textit{notify\_batch\_end}() & check the need of scaling \\
			\bottomrule
		\end{tabular}
	}	
\end{table}

\subsection{API for Cluster Scheduler}	\label{api:scheduler}

We assume that there is a centralized cluster scheduler (e.g., YARN), which has knowledge of resource availability and job status to make scheduling decisions. The scheduler may instruct EDL to adjust the parallelism of a job, identified by a unique \textit{job\_handle}, using the \textit{sclae\_in}() and \textit{sclae\_out}() operators. When a scaling operator is called, a message is sent to the leader of the workers that execute the job. The leader then coordinates the removal/addition of the specified GPU(s) and replies an acknowledgment message to the scheduler after the adjustment completes. Scaling operations are committed sequentially in EDL and if a scaling request is received in the middle of a parallelism adjustment, the leader  sends a retry message to the scheduler. The leader may fail to reply in case of failure (either the leader itself or a worker). In either case, the scheduler may retry the scaling operation after a specified time (e.g., 60 seconds). The \textit{profile}() operator measures the throughput and GPU efficiency of a job under a range of parallelism specified by [\textit{min\_p, max\_p}]. It can be used to find the optimal parallelism of a job or collect information for scheduling by running profiling tasks on a dedicated small cluster~\cite{PengBCWG18eurosys, GuCSZJQLG19nsdi}. For a job that is already running, \textit{profile}() can be called to report its throughput and GPU efficiency under the current parallelism without specifying the range.

EDL automatically recovers a job from failure using the remaining resources without intervention from the scheduler~(\S\ref{design:efficiency}), which eliminates delays due to re-scheduling and re-launching. When applying scaling, EDL keeps the aggregate batch size of all the workers constant and decides the per-worker batch size according to the parallelism. Moreover, EDL ensures that training goes over the dataset once in each epoch without repetition and omission. The above consistency semantics are sufficient for most DNN training jobs~\cite{HaochenS19icml, GoyalDGNWKTJHcorr17, bottou2009curiously}.

\subsection{API for DL Frameworks}	\label{api:framework}

EDL provides a simple API for users of popular DL frameworks to run their scripts as elastic jobs. Some of them are standard and similar to the ones in Horovod, e.g., \textit{init}(), \textit{shutdown}() and \textit{all\_reduce}(), while \textit{elastic\_shard\_generator}() and \textit{notify\_batch\_end}() are specifically introduced for elasticity. \textit{elastic\_shard\_generator}() returns a generator object, which gives the meta-data of a chunk of training samples to a worker when its \textit{next}() method is called, and a DL framework can use it to load training samples dynamically from a list of partitions. This operator ensures the efficient distribution of the training samples to a dynamic set of workers under scaling~(\S\ref{design:correctness}). EDL adds/removes workers for a job at the end of a mini-batch so that no training progress is lost and users can call \textit{notify\_batch\_end}() to notify EDL of the mini-batch boundary. The end of  a mini-batch can be identified trivially in users' training script in existing DL frameworks, for example, after \textit{session.run()} in TensorFlow and the end of the \textit{for-loop} for each mini-batch in PyTorch. Since each mini-batch typically takes hundreds of milliseconds, the delay of waiting for the end of a mini-batch is usually short. %


\begin{lstlisting}[label=tf-sample-code, caption=An example code of EDL with TensorFlow, captionpos=b, float=tp]
import tensorflow as tf
import edl.tensorflow as edl
edl.init() # initialize EDL daemon
# create a generator object
ds = tf.data.Dataset.from_generator(
edl.elastic_shard_generator())
loss = Resnet50(ds) # construct a Resnet50 model
# ingest Allreduce into graph within Optimizer
opt = edl.Optimizer(tf.train.AdamOptimizer(...))
# create optmization objective
obj = opt.minimize(loss) 

with tf.train.Session() as s:
while not s.Done():
s.run(obj, feed_dict={...})
edl.notify_batch_end()
\end{lstlisting}

Putting things together, we illustrate with an example that uses EDL with TensorFlow in Listing~\ref{tf-sample-code}. Line~3 initializes the EDL daemon and Lines~5-6 construct a TensorFlow dataset object from the \textit{elastic\_shard\_generator}() method of EDL. The \textit{edl.Optimizer} in Line~9 is a helper class that inherits TensorFlow's optimizer class and we ingest the Allreduce operation into the computation graph in \textit{edl.Optimizer}. In Line~16, users indicate the end of one mini-batch with \textit{notify\_batch\_end}(). It can be seen that using EDL is easy and it only requires adding a few lines (i.e., the lines containing ``edl'') to a user's script.

\section{System Design and Implementation}\label{sec:design}

The design of the EDL system has three goals: \textit{flexibility}, \textit{efficiency}, and \textit{consistency}. Flexibility means that EDL should allow any process, either a worker or the leader, to leave or join a job at any time, which enables the scheduler to flexibly adjust parallelism. Efficiency means that EDL should significantly reduce the parallelism adjustment overhead compared with stop-resume, and should introduce negligible overheads to training under static parallelism without scaling. Consistency means that the execution of a job under scaling should be equivalent to its execution without scaling~\cite{HaochenS19icml, GoyalDGNWKTJHcorr17, bottou2009curiously}.

To achieve these goals, EDL adopts three key designs: \textit{automatic job management}~(\S\ref{design:flexibility}), \textit{efficient parallelism adjustment}~(\S\ref{design:efficiency}), and \textit{dynamic data pipeline}~(\S\ref{design:correctness}).

\subsection{Automatic Job Management}\label{design:flexibility}

Each job has a leader to manage its workers. However, the leader may leave the job due to scaling or failure. Each worker runs a leader election/discovery procedure whenever the leader is not known to the worker, which ensures that there is always a leader to manage the job. Specifically, when a job is launched, each worker first performs the leader election procedure, which is implemented as a distributed \textit{compare\_and\_swap} transaction using an external coordination system such as ZooKeeper~\cite{HuntKJR10atc} or etcd~\cite{etcd}. The workers query the leader's connection information (e.g., hostname and port number) in the external coordination service using the \textit{job\_handle} as key. If the connection information is void or expired, a worker writes its own address into the information and becomes the leader. The leader needs to periodically refresh its address information, which is configured to expire automatically if the leader fails to do so. Upon expiration, workers will be notified so that they will perform leader election again.

After a leader is elected, it establishes an RPC server accepting connections, while other workers connect to the leader and send a registration message to join the job. According to our measurement, leader election took 7ms on average and 33ms at maximum when 256 workers used etcd for distributed coordination. During job execution, the leader infers the liveness of the workers from the gradient synchronization requests in every mini-batch and thus explicit heartbeat message is not needed. When  \textit{sclae\_out}() or \textit{sclae\_in}() is called, the leader communicates with the new or exiting workers to prepare them for joining or leaving the job. The leader also constructs a new communication topology for distributed training with/without the new/exiting workers. More details will be presented when we discuss  \textit{sclae\_out}() and \textit{sclae\_in}() in~\S\ref{design:efficiency}.

An alternative to the leader discovery mechanism is to launch a dedicated process (not attached with GPU) as the leader (similar to an application master~\cite{VavilapalliMDAKEGLSSSCORRB13cloud}). Such a design has the advantage that \textit{sclae\_in}() operations will not affect the leader. However, using multiple types of processes complicates the current single-program-multiple-data (SPMD) execution pattern. Deployment is also more complicated as users need to configure parameters such as the network address and resource requirements for the leader. 

\subsection{Efficient Parallelism Adjustment} \label{design:efficiency}

\begin{figure}
	\centering
	\includegraphics[width=0.35\textwidth]{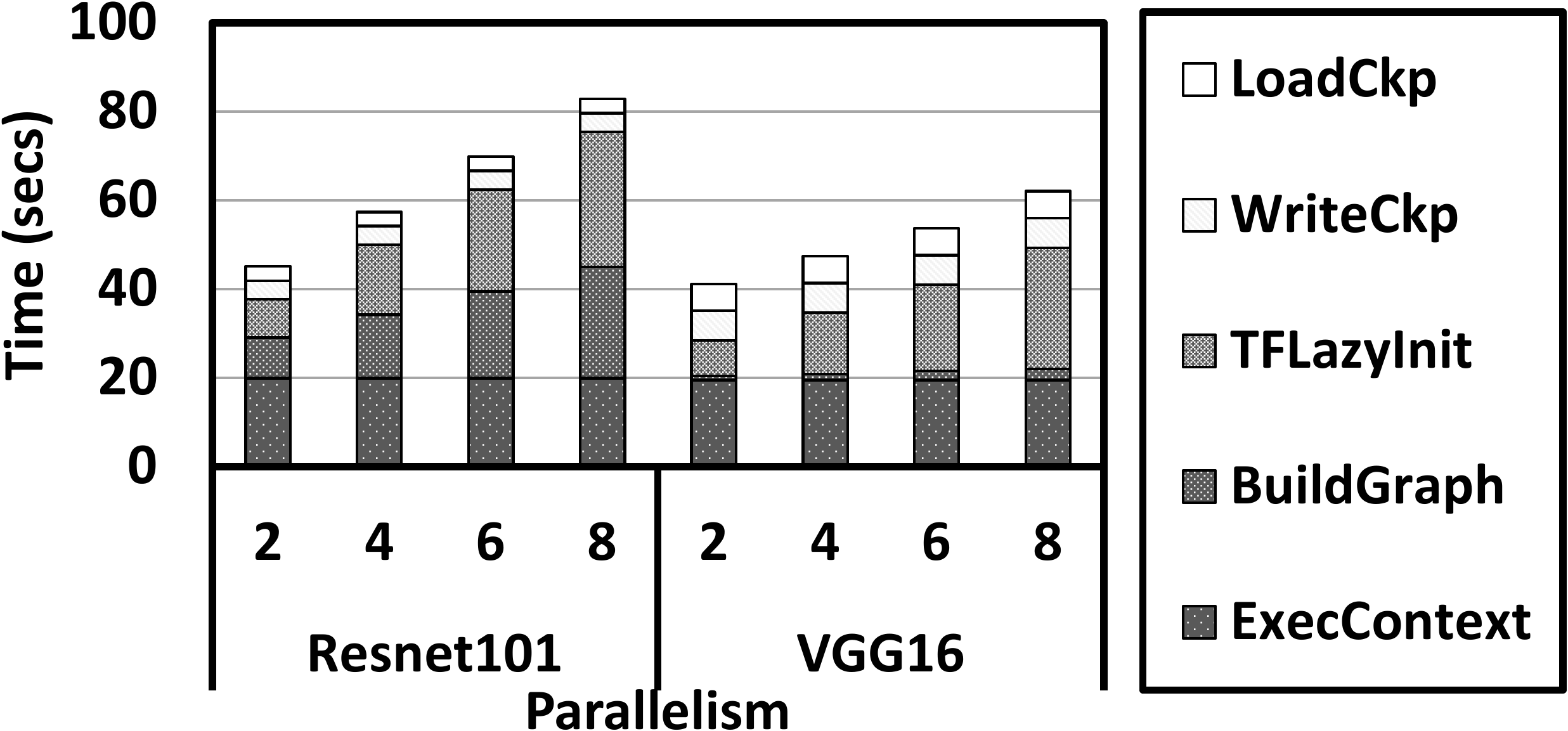}
	\caption{Scaling overhead decomposition for TensorFlow}
	\label{fig:scaling}
\end{figure}

To reduce the overheads of parallelism adjustments, EDL uses \textit{stop-free scaling} to hide the high cost of \textit{execution context preparation} during  \textit{sclae\_out}() and applies \textit{graceful exit} to make the overhead of \textit{sclae\_in}() negligible. 

\paragraph{Scale out.} \ Adding new workers to a running job takes three steps: \textit{execution context preparation}, \textit{communication topology construction}, and \textit{model preparation}. Execution context preparation involves loading dynamic libraries (e.g., cuDNN, cuBLAS), preparing training data, allocating space on both GPU memory and main memory, and so on. Declarative DL frameworks such as TensorFlow also need to build and optimize the computation graph. For communication, new workers need to connect to the leader for coordination and all the workers need to form a new ring topology for model synchronization. New workers also need to acquire the up-to-date model before joining the training. We provide a breakdown of the time for scaling out a 1-GPU job on TensorFlow in Figure~\ref{fig:scaling} with the execution context preparation overhead marked in gray. The result shows that the cost of execution context preparation dominates the scaling overhead. This observation is consistent for all the models we experimented.

Motivated by this observation, we propose~\textit{stop-free scaling}. The key insight is that the training on the existing workers does not need to be stopped when the new workers conduct execution context preparation. Each new worker launches two separate threads, a main thread and a background thread. The main thread conducts execution context preparation while at the same time the background thread performs leader discovery and sends a registration request to the leader. The leader constructs a new communication topology involving the new workers after receiving their registration requests and broadcasts it to all the workers. Note that the original communication topology is not destructed yet, and thus the existing workers can continue the training without being affected. A new worker sends a ready message to the leader when it finishes execution context preparation and receives the new communication topology, but is blocked until it receives an OK message from the leader. 

\begin{figure}[!t]	
	\centering 
	\includegraphics[width=0.48\textwidth]{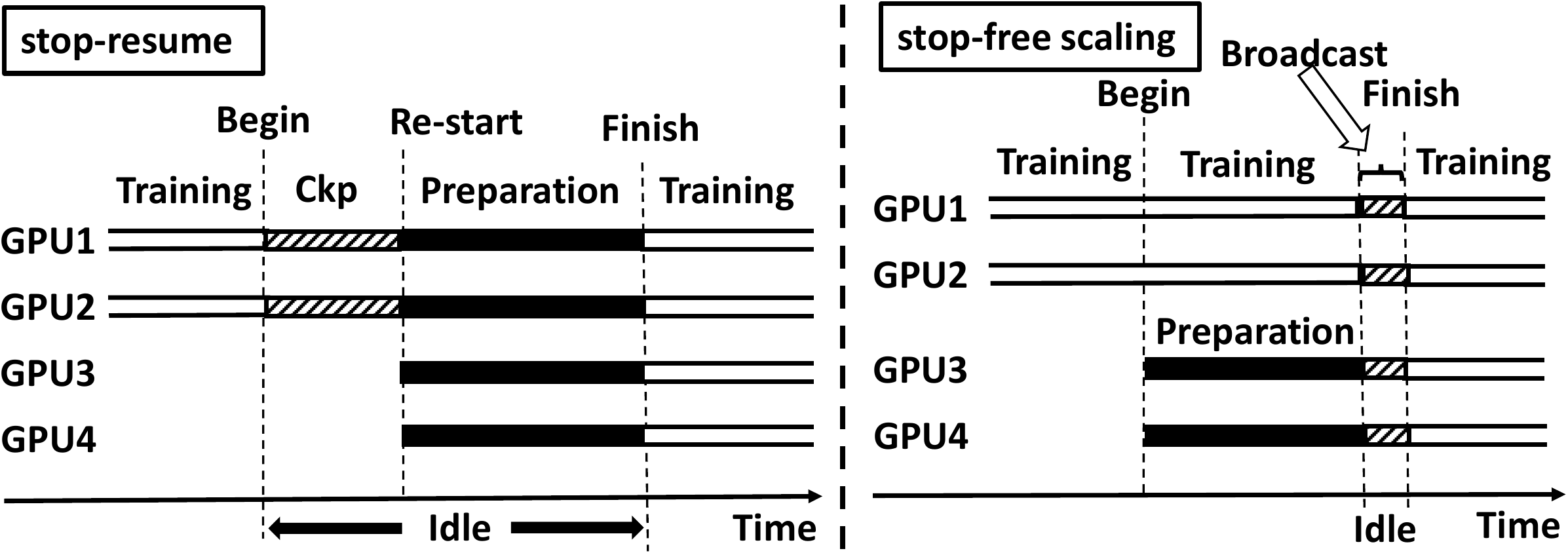}
	\caption{An illustration of stop-free scaling}
	\label{fig:stop-free}
\end{figure}                    

Once the ready messages from all the new workers have been received, the leader broadcasts an OK message and a future timestamp to all the workers. The existing workers check at the end of each mini-batch indicated by \textit{notifyBatchEnd}() and switch to the new communication topology when its next local timestamp reaches the timestamp specified by the leader. The timestamp is implemented as the mini-batch count and we set the future timestamp as $t\_cur + k$, where $t\_cur$ is the current mini-batch count of the leader. $k$ is determined as $T_a/T_b$, in which $T_b$ is the current per-mini-batch time for the job and $T_a$ is a predefined time allowance (500ms by default) to tolerate fluctuations in network latency. One existing worker is chosen to broadcast its model to the new workers as using only one worker for broadcasting reduces the time for model synchronization. After the new workers obtain the latest model, \textit{sclae\_out}() completes and the training continues with the new parallelism.

An illustration of stop-free scaling, contrasting with stop-resume,  is given in Figure~\ref{fig:stop-free}, where we add two more GPUs to a job. With stop-free scaling, existing workers only need to stop and wait until the model is broadcast to the new workers, which can complete within 1 second for most models according to our experiments. Compared with stop-resume, the long execution context preparation time for new workers is now hidden behind the normal execution of the existing workers. 

\paragraph{Scale in.} \ For \textit{sclae\_in}(), we apply \textit{graceful exit}, in which the scheduler gives the exiting workers a short time allowance (e.g., 30 seconds, but usually a few seconds is enough) to leave. On receiving the \textit{sclae\_in}() request, the leader constructs a new communication topology and broadcasts it to the remaining workers. Similar to the case of \textit{sclae\_out}(), the leader also sends a future timestamp to all the workers, at which the exiting workers should leave and the remaining workers should switch to the new communication topology. Before reaching this timestamp, training continues with all the workers. If the leader is instructed to leave, it will erase its address in the external coordination system such that a new leader can be elected using the leader election protocol. The old leader will send the job meta-data (e.g., batch size, data loading progress, etc.) to the new leader before exiting and all the remaining workers will connect to the new leader at the scheduled timestamp. With graceful exit, the overhead of \textit{sclae\_in}() is negligible as the exiting workers just need to leave and the remaining workers do not need to stop and wait.

\paragraph{Failure recovery.} \ We consider \textit{forced exit}, including process failure, as a special case of scaling in. Worker failure can be detected if a worker fails to send the gradient synchronization request for a mini-batch and leader failure can be detected by the leader election/discovery protocol. When failure happens, the model may be inconsistent. For example, if a worker fails before finishing synchronizing all gradients, the model on the other workers would be partially updated. EDL provides two protocols to recover from failure, i.e., ~\textit{consistent recovery} and~\textit{approximate recovery}. Consistent recovery requires the leader to write a checkpoint to persistent storage such as HDFS~\cite{ShvachkoKRC10mss} periodically (e.g., every 1000 mini-batch or every 10 minutes). Upon failure, the job is resumed by loading and restarting from the latest checkpoint, which ensures model consistency. As DNN training is known to be robust to bounded errors, approximate recovery can also be used to simply construct a new communication topology for the surviving workers and redo the current mini-batch. Users can choose one of the two protocols by altering the value of the environment variable USE\_APPX\_RECOVERY (our default is using consistent recovery).

\subsection{Dynamic Data Pipeline}\label{design:correctness}

Existing DL frameworks partition a dataset among workers before training starts, and each worker goes over its assigned partitions in each epoch~\cite{AbadiBCCDDDGIIK16osdi, Paszke2017automatic, ChenLLLWWXXZZ15corr}. This \textit{static data allocation} method works well in practice, but we show that static data allocation lacks flexibility and results in complicated data management for elastic DL. 

Consider a dataset with 1M samples, which is partitioned into 1K partitions each with 1K samples, and there are 10 workers each getting 100 partitions. If we want to add 5 GPUs to this job, two options are possible under static data allocation. First, we can wait until the end of the current epoch and re-assign the partitions among the 15 workers, which is inflexible as parallelism adjustment is only possible at the end of the current epoch (instead of the current mini-batch as in EDL). Second, we can re-assign only those unprocessed partitions in the current epoch among the 15 workers and conduct a global re-allocation when the current epoch ends. However, if another scaling instruction (e.g., removing 3 out of the 5 added GPUs because they are transient resources) comes before the re-assignment finishes, a new data allocation plan needs to be constructed on the partially re-assigned data within the current epoch. Some other issues, such as hiding the delay of data re-assignment and handling partition fragmentation or imbalance, also need to be considered, which make the design and implementation complicated.

To support elasticity, EDL assigns data partitions to workers dynamically in an on-demand fashion. The dataset is logically divided into $d$ partitions, where $d$ is sufficiently larger than the number of workers while the size of a partition is still large enough to allow high-bandwidth data reading. The partitioning is only conducted at the meta-data level, e.g., recording file names and offsets, and the dataset is not physically partitioned. The leader generates a random permutation of the indexes of the partitions and uses it for dynamic data assignment. When a worker needs a new partition, it sends a data-read request to the leader by calling the \textit{next}() method of the generator object returned by \textit{elastic\_shard\_generator}(). The leader replies the request with the meta-data (e.g., file path, offset and length) of the next unassigned partition. The worker then issues asynchronous I/O request to the distributed file system (e.g., HDFS~\cite{ShvachkoKRC10mss}) for reading this partition.

For the purpose of progress tracking, each worker records an offset in its current patition, which indicates where the next mini-batch should start. The workers report their offsets to the leader at the end of each mini-batch and this information is attached to the gradient synchronization request with negligible overhead. When new workers join a job, the leader simply assigns some unprocessed (or partially processed) partitions to them. When a worker leaves under graceful exit, it reports to the leader the meta-data of the current partition and its offset in the partition such that the leader can assign the remaining unprocessed data in this partition to another worker. If the leader needs to leave, it sends the partition permutation list and the progress of all the workers to the new leader before it exits. EDL also writes the partition permutation list and the worker progresses to checkpoint such that a job can be restored properly. 

The above procedure of dynamic data pipeline in EDL ensures that training goes over the dataset once in each epoch without repetition and omission regardless of whether scaling out and in are performed. However, different runs of an algorithm may not produce the same result as scaling may affect the order in which the samples are used in the training. In essence, the change in the processing order of the samples caused by scaling can be viewed as an additional source of randomness in the sample permutation and thus the consistency guarantee by dynamic data pipeline is sufficient for most deep learning tasks~\cite{HaochenS19icml, GoyalDGNWKTJHcorr17, bottou2009curiously}.

\subsection{Implementation Details}

We modified Horovod v0.16.1 and implemented the EDL daemon and plugins using Boost.asio with around 4K lines of code. We use NCCL v2.4.8 and TensorFlow v1.14.1. TCP is used to connect the leader with the workers and the cluster manager. We observed that usually tens of coordination messages are exchanged between the leader and the workers in each mini-batch training and the size of each message is within a few hundred bytes. As each mini-batch training usually takes only a few hundred of milliseconds, reducing the messaging latency is critical to avoid wasting GPU cycles. Therefore, we disabled the Nagle's algorithm~\cite{MurrayMIIBA13sosp, Naglerfc896} in the TCP socket and the average latency of sending one message is 56 $\mu s$ according to our measurement. We are also investigating to use RDMA to further reduce the latency.

To hide the latency of reading training data from the file system, each worker runs a producer-consumer data pipeline. A ping-pong buffer (or double buffer) is maintained between CPU and GPUs. The buffers are blocks of pinned memory to avoid disk swapping and enable fast data transfer to GPUs. A background thread serves as the producer and asks the leader for the meta-data of a new partition once a partition is dequeued from one of the buffers by the consumer. We overlapped host-to-device data movement with GPU computation by pre-fetching multiple mini-batches of training samples from main memory to GPU.

\section{Elasticity In Use}\label{sec:usecase}

In this section, we discuss how EDL can benefit DL cluster scheduling and be used to implement a number of important system functionalities such as straggler handling, performance profiling, and worker migration.

\subsection{Elasticity-Aware DL Scheduling}	\label{usecase:scheduling}

According to~\S\ref{sec:background and motivation}, EDL can be used to (1)~adjust the trade-off between throughput and GPU efficiency, (2)~improve cluster utilization and JCT by adapting to the variations in cluster load, and (3)~make good use of transient idle resources. One way to enjoy all of these three benefits is an elasticity-aware DL scheduler based on EDL. 


As developing a new scheduler is out of the scope of this paper, we extend Tiresias~\cite{GuCSZJQLG19nsdi}, a state-of-the-art GPU cluster scheduler  based on the shortest-job-first principle. Tiresias manages jobs in multiple groups, $G_0, G_1, \dots$, and the group with a smaller index has higher priority. Scheduling is conducted by allocating resources to jobs in the higher-priority groups first. Each group $G_i$ has a service quantum $t_i$ for its jobs, meaning that a job can only consume up to $t_i$ GPU*sec and after that it will be moved to $G_{i+1}$. When a job is submitted to the cluster, it is first placed into $G_0$ and gradually moved to a lower-priority group as it keeps running. If a job is not scheduled for a long time, it will be moved to $G_0$ to prevent starvation. Tiresias computes a new scheduling plan for all jobs whenever there is a new event (e.g., a new job is received or some job changes its priority). A running job will be preempted if it cannot be scheduled (i.e., its required resources cannot be allocated) in the new plan. Tiresias achieves good responsiveness for small jobs because they can be completed in the first few groups, i.e., groups with higher priority (e.g., jobs that take less than $t_0$ GPU*sec are always scheduled first).  Readers may refer to~\cite{GuCSZJQLG19nsdi} for details.

To extend Tiresias to support elasticity, we add the following two simple rules to its scheduling protocol. 
\begin{itemize}
	\item \textbf{(R1: Compaction)}~If the number of waiting jobs (waiting to be scheduled) exceeds a threshold $N$, we scan the pending jobs starting from highest priority group. For each pending job $\tilde{J}_i$, we calculate $Gain(i,p)$, which is the gain in GPU efficiency by removing $p$ GPUs from the running jobs via scaling in and allocating these GPUs to $\tilde{J}_i$. We find the $p$ that maximizes $Gain(i,p)$ as the scheduling plan~\footnote{We enforce a locality constraint that the $p$ GPUs to be allocated to $\tilde{J}_i$ must come from no more than $\lceil p_i/m \rceil$ machines, where $p_i$ is the user-specified parallelism for $\tilde{J}_i$ and $m$ is the number of GPUs on each machine.}. We do not remove GPU from jobs in $G_0$ and constrain that a running job $J_j$ has a parallelism at least $\lceil rp_j \rceil$ to guarantee quality of service, where $p_j$ is the user-specified parallelism for $J_j$ and $r \in (0,1]$. 
	\item \textbf{(R2: Expansion)}~If there is no waiting job and there are idle GPUs, \textit{sclae\_out}() is to be applied to jobs in a greedy manner as follows. In each step, the job that has the largest gain is to be allocated 1 more GPU, where the gain is defined as $\frac{S(p+1)-S(p)}{S(p)}$, in which $S(p)$ is the training throughput with the current parallelism $p$. The greedy procedure continues until all the idle GPUs are allocated or no job can obtain a positive gain. 
	
\end{itemize}

We call the new scheduling algorithm \textbf{Elastic-Tiresias}. Intuitively, the two rules of Elastic-Tiresias aim to improve GPU efficiency when the cluster load is high and try to fully utilize the idle resources when the load is low. With these two simple modifications, Elastic-Tiresias achieves significantly better performance compared with the original Tiresias~(\S\ref{exp:scheduling}). If users do not want the scheduler to change the parallelism of a job, they can mark the job as inelastic and Elastic-Tiresias simply skips it when conducting parallelism adjustment.   

\subsection{Additional Use Cases of EDL}\label{usecase:others}

EDL can also be easily used to provide important system functionalities such as follows.

\paragraph{Straggler mitigation.}  Some workers may become stragglers due to reasons such as high GPU temperature (which leads to clock frequency drop) and strong interference from co-located jobs. Stragglers are a major cause of performance degradation in synchronous training as a synchronization barrier is enforced at the end of each mini-batch. EDL detects stragglers by monitoring the time workers spend on a mini-batch via the gradient synchronization requests. If a worker is consistently slower than other workers in a few consecutive mini-batches (e.g., its per-mini-batch time is longer than 1.2 times of the median for 10 mini-batches), the leader may trigger a \textit{sclae\_in}() operation to remove this worker from training with negligible overhead. Note that a smaller parallelism without straggler can lead to better performance as we report in~\S\ref{exp:straggler}. A replacement worker, or the straggler itself (e.g., after cooling down or the completion of co-located jobs), can easily join the job using \textit{sclae\_out}() to restore to the original parallelism.

\paragraph{Performance profiling.} Building an analytical model for the performance of DNN training jobs under different parallelism and placement plans is important but generally challenging~\cite{PengBCWG18eurosys, GuCSZJQLG19nsdi, XiaoBRSKHPPZZYZ18osdi}. There are many factors such as model architecture, global batch size and network bandwidth, that may affect performance in different ways. Therefore, it is common and also often necessary to run profiling jobs to measure the performance under different configurations to collect information for performance tuning and/or job scheduling. The \textit{profile}() method in EDL can be easily used to measure the runtime performance under a range of parallelism, defined by $[\text{min}, \text{max}]$. As \textit{sclae\_in}() has much lower overhead than \textit{sclae\_out}(), EDL starts a profiling job with the maximum parallelism and gradually scales in to the minimum parallelism. At each parallelism, the job is run for a few mini-batch iterations (e.g., 20) to measure the performance.

\paragraph{Worker migration.} Sometimes the scheduler needs to move one worker of a job from one machine to another machine, e.g., to co-locate the workers of this job for communication cost reduction or to make room in a machine so that it can be dedicated for some purposes. Worker migration can be easily operated in EDL by first scaling in to remove the workers on the destination machine and then scaling out to add new workers from the target machine, without stopping the job. We further optimize this procedure by merging the scale-in and scale-out operations into one single migration operation, in which the communication topology is switched only once.     

\section{Experimental Results} \label{sec:experiment}

We evaluated EDL on a cluster with 8 machines each with a 96-core Intel CPU, 8 NVIDIA Tesla V100 SMX2 GPUs and 256 GB RAM. The machines are connected with 100 Gbps infiniband.

\subsection{The Overheads of Elasticity}	\label{exp:overhead}

\begin{figure}[!t]	
	\centering 
	\begin{minipage}[b]{0.23\textwidth}
		\centering
		\includegraphics[width=\textwidth]{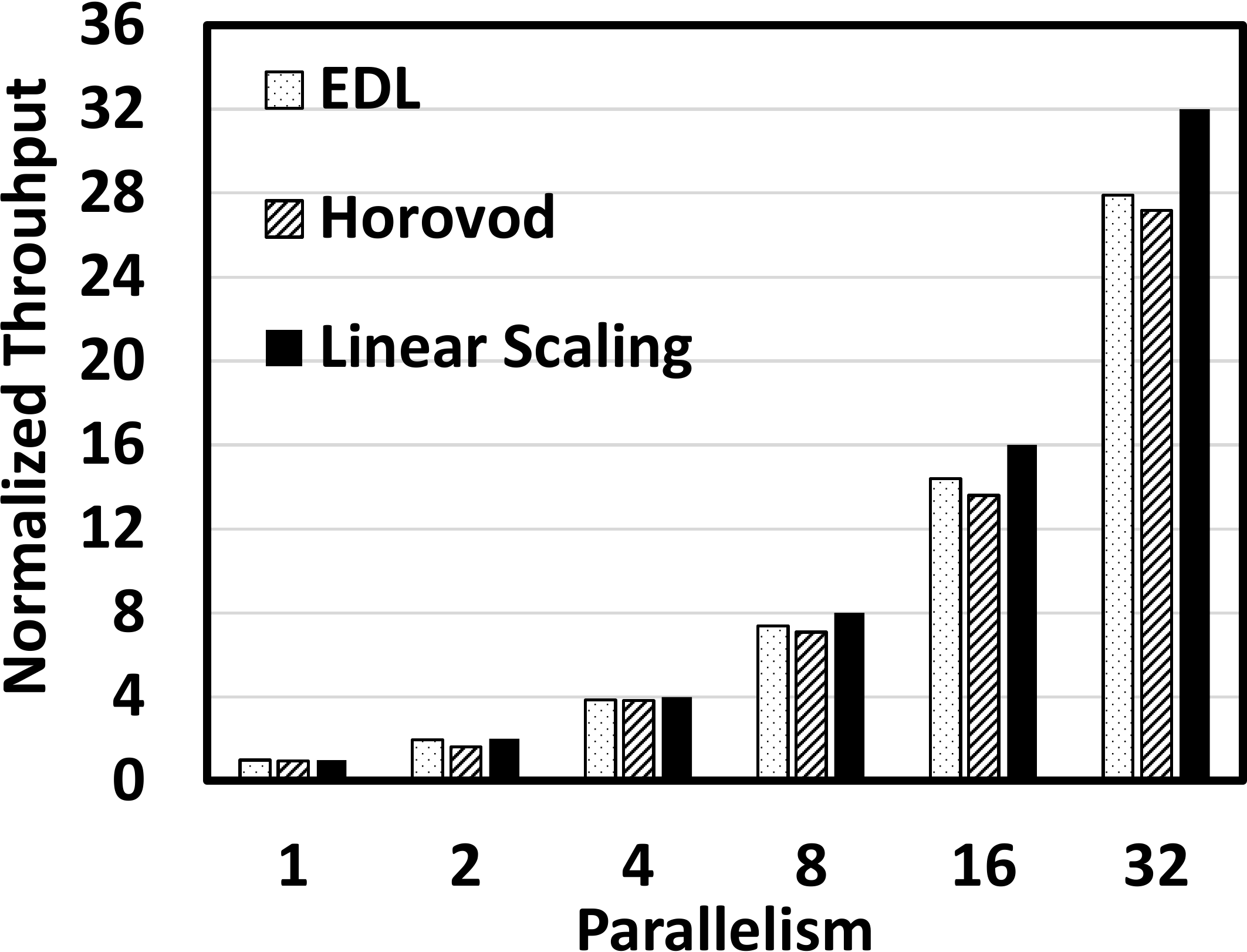}
		\subcaption{ResNet101}
	\end{minipage}
	\begin{minipage}[b]{0.23\textwidth}
		\centering
		\includegraphics[width=\textwidth]{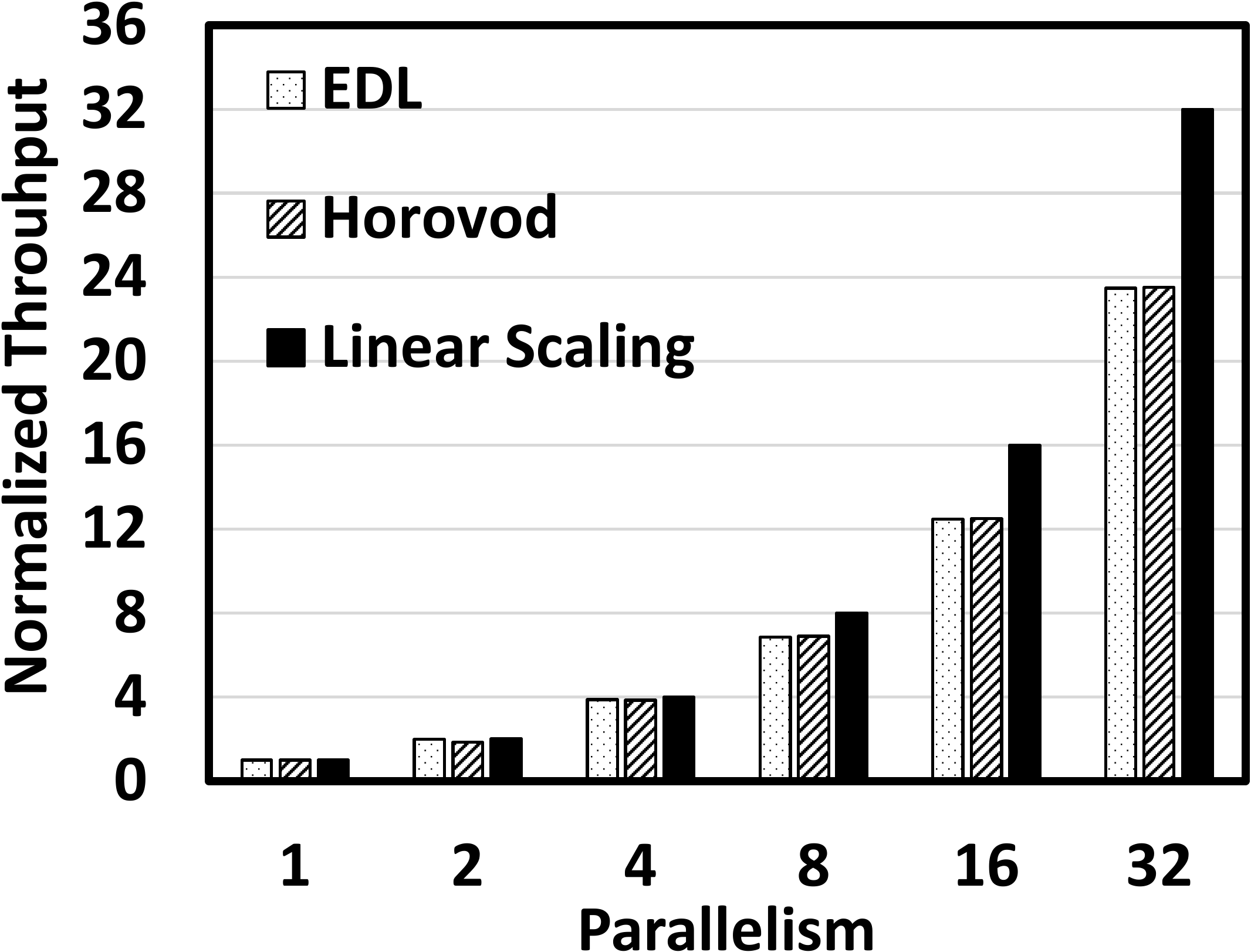}
		\subcaption{VGG16}
	\end{minipage}
	\vspace{-2mm}
	\caption{Performance under static parallelism}
	\label{fig:static parallelism}
\end{figure}

\paragraph{Performance under static parallelism.} As DNN training jobs run with a static parallelism most of the time, it is crucial that the designs for elasticity, e.g., RPC-based coordination and dynamic data pipeline, incur little overhead on normal training. We measured the throughput (averaged over 500 mini-batches) of EDL and Horovod for training different DNN models using up to 32  GPUs. As a common practice of testing the scalability of distributed DNN training systems, we increased the total batch size linearly with the number of GPUs. Due to the page limit, we only report the results for ResNet101 and VGG16 in Figure~\ref{fig:static parallelism}, which show that EDL achieves comparable performance with Horovod and scales almost linearly for ResNet101. The scalability of both systems for training VGG16 drops a bit due to the high communication cost caused by the large model.



\begin{table}[t]
	\centering
	\caption{Stopping time (sec) of scaling out}
	\label{tab:scaling time}
	\vspace{-3mm}
	\begin{center}
		\fontsize{8}{9}\selectfont
		\begin{tabular}{cccccl}
			\toprule		
			&  AlexNet & ResNet152 & ResNet50 & VGG19 & VGG16  \\
			\midrule
			Stop-resume & 30 & 70       &44         & 38  & 35 \\
			EDL &0.18& 1.8      & 0.67      & 0.71 & 0.36  \\
			\bottomrule
		\end{tabular}
	\end{center}
\end{table}

\begin{table}[t]
	\centering
	\caption{End-to-end time (sec) of scaling in/out in EDL}
	\label{tab:end-to-end time}
	\vspace{-3mm}
	\begin{center}
		\fontsize{8}{9}\selectfont
		\begin{tabular}{cccccl}
			\toprule		
			&  AlexNet & ResNet152 & ResNet50 & VGG19 & VGG16  \\
			\midrule
			Scaling in & 1.6 & 3.3       & 1.8        & 3.3  & 3.3 \\
			Scaling out & 16 & 36     & 21        & 20  & 19 \\
			\bottomrule
		\end{tabular}
	\end{center}
\end{table}

\paragraph{Scaling overheads.} To scale in, EDL does not stop training and uses graceful exit to remove exiting worker(s). To scale out, EDL needs to stop training briefly to broadcast the model to new worker(s)~(\S\ref{design:efficiency}). In comparison, stop-resume needs to stop all workers for the entire period for both scaling in and out. We report the stopping time (averaged over 20 trails) of scaling out from 4 GPUs to 5 GPUs with EDL and stop-resume in Table~\ref{tab:scaling time}. Note that stop-resume has similar stopping time for scaling in. Other factors, e.g., scaling to more GPUs or the location (on the same machine or another machine) of the added GPUs, have negligible influence on the scaling overhead and thus we omit the details.  

We also report the end-to-end time of scaling in (from 5 to 4 GPUs) and scaling out (from 4 to 5 GPUs) operations in EDL in Table~\ref{tab:end-to-end time}. We remark that (1)~the reported time only affects the joining/exiting workers, while normal training continues with existing workers, and (2)~scaling in/out with more GPUs does not linearly increase the end-to-end time (which is only affected by the slowest worker that completes the scaling process).  The results show that the scaling operations in EDL can finish quickly.

\begin{figure}[!t]	
	\centering 
	\includegraphics[width=0.23\textwidth]{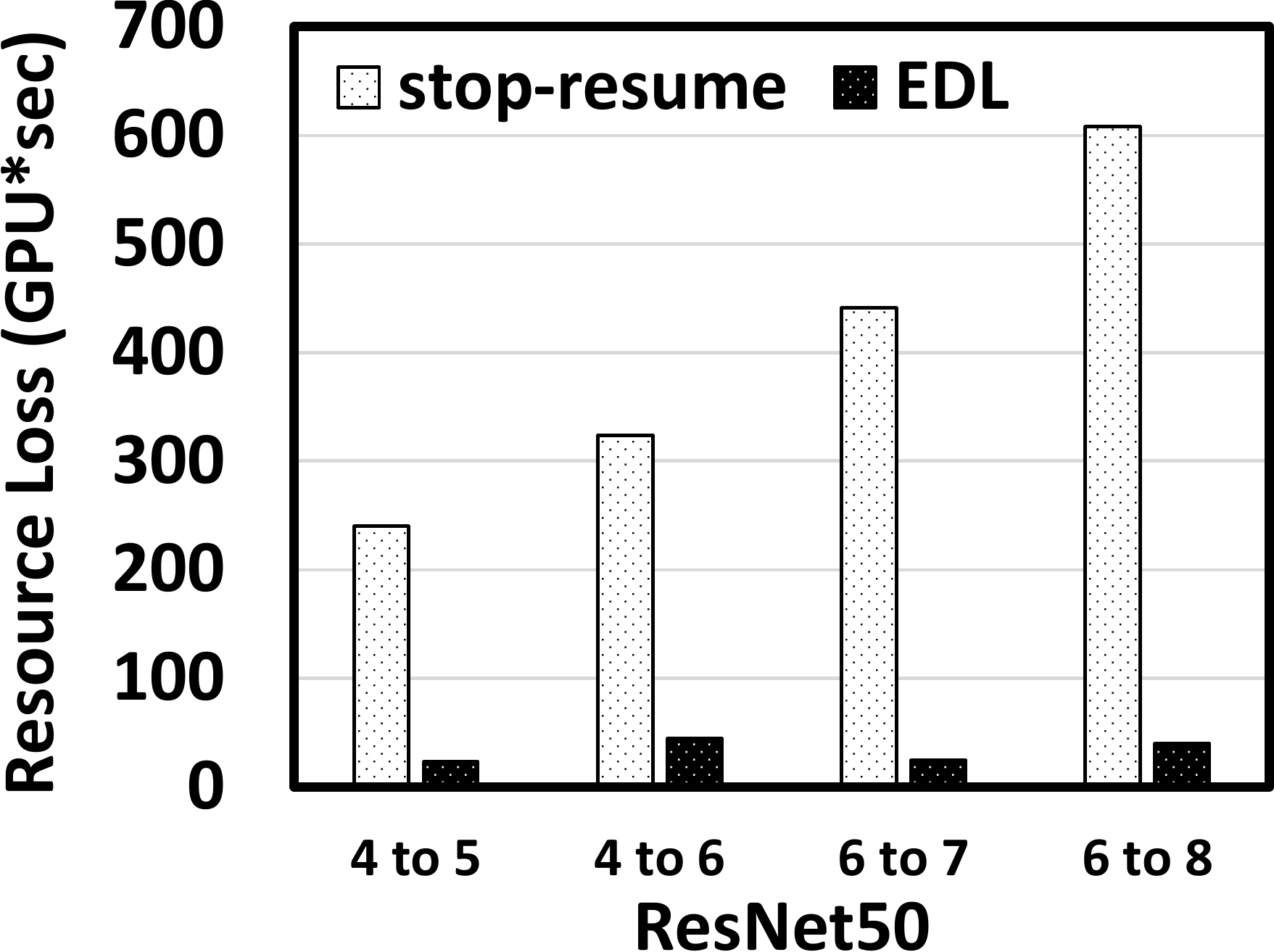}
	\includegraphics[width=0.23\textwidth]{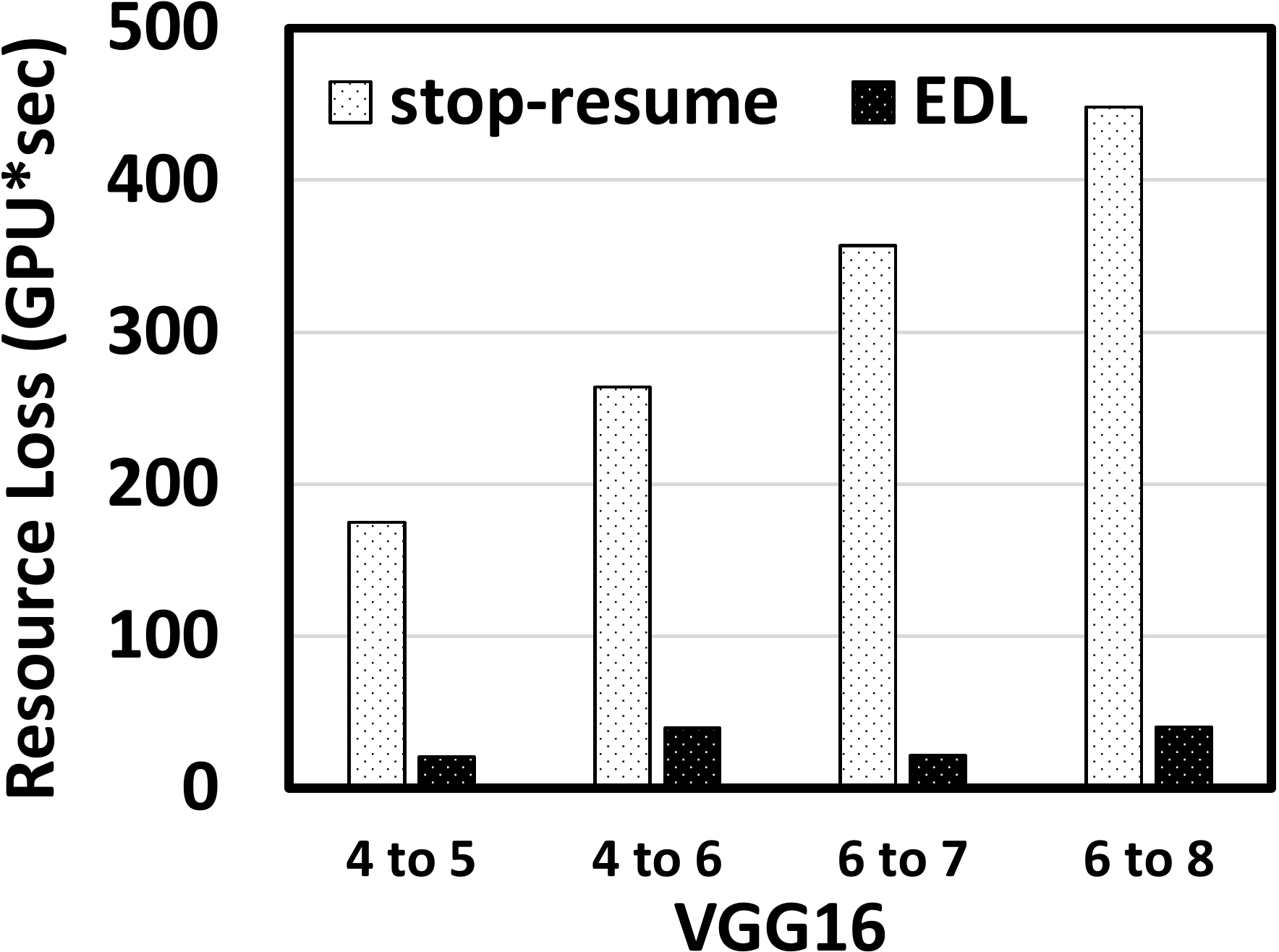}
	\vspace{-2mm}	
	\caption{GPU resource loss of scaling out}
	\label{fig:loss}
\end{figure}

As the longer the stopping time ($T_s$) and the end-to-end time ($T$), the more is the GPU resource not used for training. Thus, we also measured the total amount of resource loss (in GPU * time) due to scaling out. We report the results for ResNet50 and VGG16 in Figure~\ref{fig:loss}, but similar patterns are observed for other models as well. The resource loss of EDL is an order of magnitude smaller because only the newly added GPUs are not used in $T$ and the existing GPUs are not used only in $T_s$; in contrast, for stop-resume all GPUs (new and old) are not used in $T$. We remark that the loss due to the new GPUs is inevitable as new workers always need to conduct context preparation before training, but this loss actually contributes to the majority of EDL's loss (as $T_s$ is small).

\subsection{The Benefits of Using EDL}	\label{exp:applications}


In this set of experiments, we demonstrate the benefits brought by EDL in various applications. 

\begin{figure}[!t]	
	\centering 
	\begin{minipage}[b]{0.23\textwidth}
		\centering
		\includegraphics[width=\textwidth]{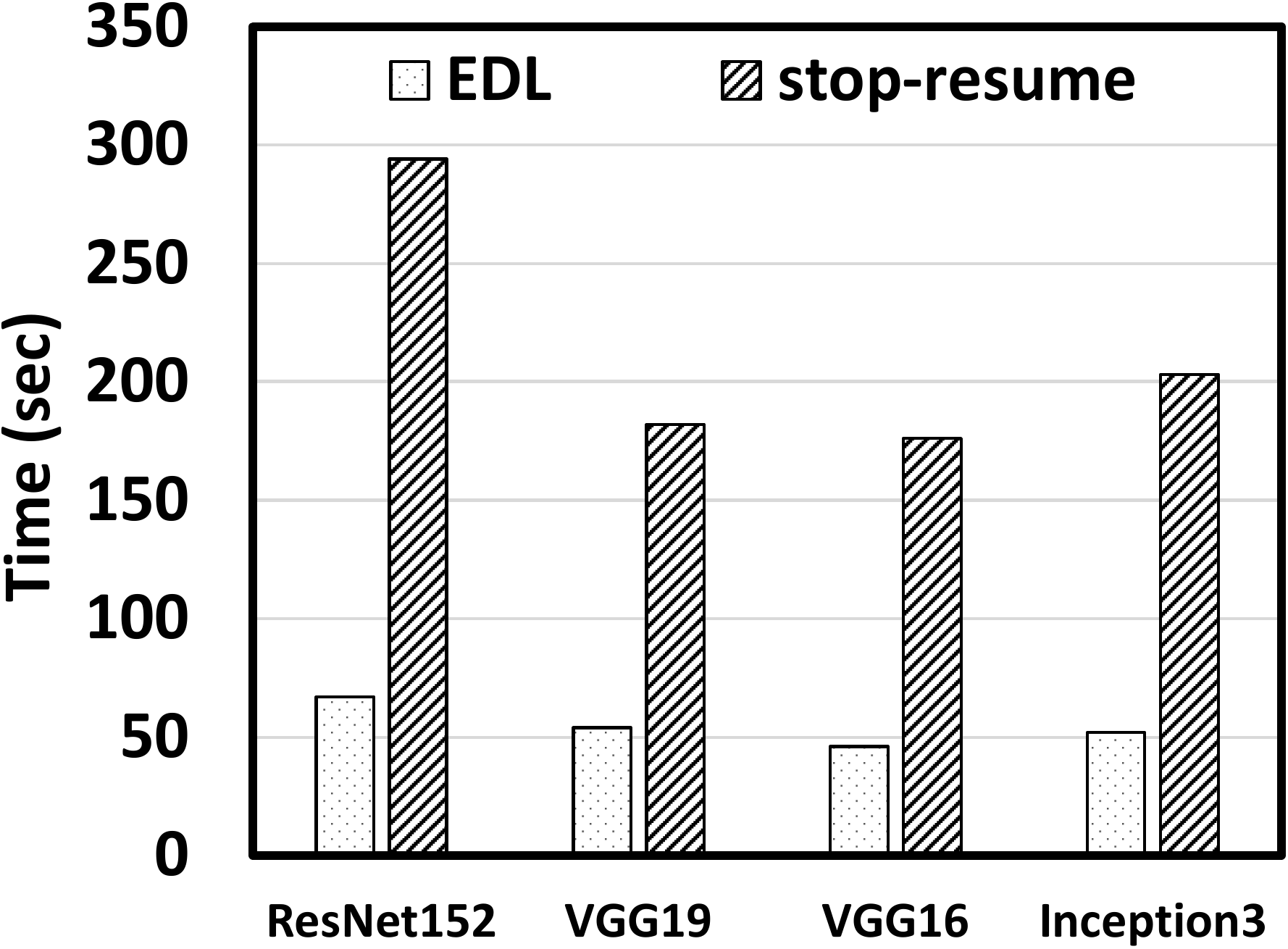}
		\subcaption{Profiling}\label{fig:profiling}
	\end{minipage}
	\begin{minipage}[b]{0.23\textwidth}
		\centering
		\includegraphics[width=\textwidth]{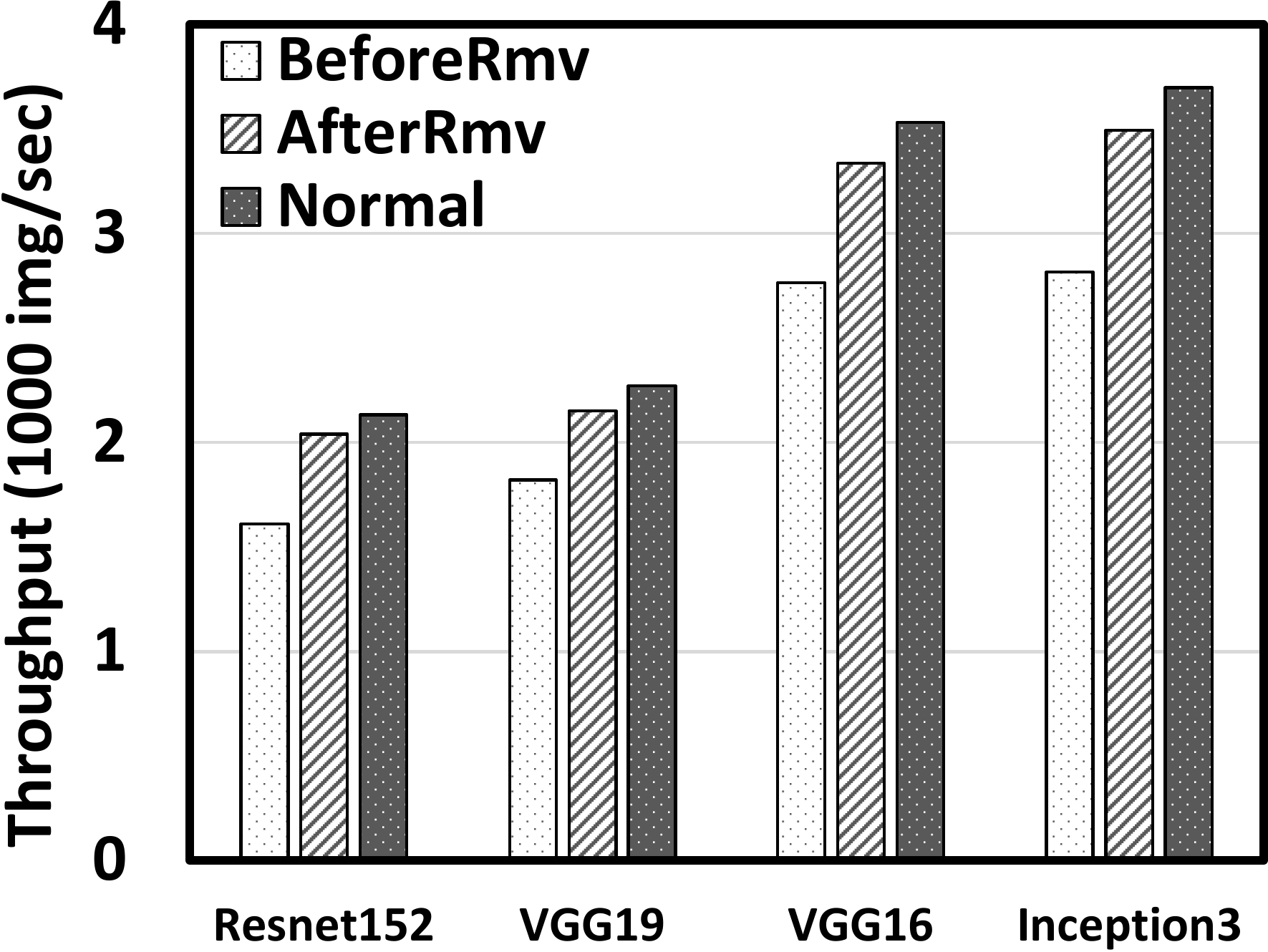}
		\subcaption{Straggler mitigation}\label{fig:straggler}
	\end{minipage}
	\vspace{-2mm}	
	\caption{Performance for profiling and straggler migration}
	\label{fig:profiling and straggler}
\end{figure}

\paragraph{Performance profiling.} We report the time taken by EDL and stop-resume for a profiling job (testing the training performance with 2 to 8 GPUs and running for 10 mini-batches under each parallelism) in Figure~\ref{fig:profiling}. EDL first started the job with 8 GPUs and then gradually scaled in to 2 GPUs. In contrast, stop-resume started a new job under each parallelism to measure the performance. The results show that EDL used approximately 20\% of the time taken by stop-resume to do the same profiling jobs. This is because stop-resume needs to pay the expensive context initialization cost repeatedly for each parallelism, while EDL pays the context initialization cost only once at the beginning and then uses low-overhead scale-in operations to adjust the parallelism.

\paragraph{Straggler mitigation.}\label{exp:straggler} We manually created a straggler for a job running with 16 GPUs, by delaying its gradient synchronization requests by 1/3 of the per-mini-batch time, which is equivalent to limiting its computation capability to 75\% of the maximum. Figure~\ref{fig:straggler} shows that the overall throughput also degrades to approximately 75\% of the normal case, as all workers need to wait for the straggler in synchronous training. We configured EDL to detect stragglers based on the statistics of the past 10 mini-batches. For all the jobs we tested, EDL took less than 10 seconds to detect the straggler and removed it within 5 seconds using scale-in. After the straggler was removed, the training throughout returned to about 94\% of the normal case (with 1 less GPU, i.e., the removed straggler). Note that when there are more stragglers, the detection time and removal time do not increase.


\begin{figure}[!t]	
	\centering 
	\begin{minipage}[b]{0.23\textwidth}
		\centering
		\includegraphics[width=\textwidth]{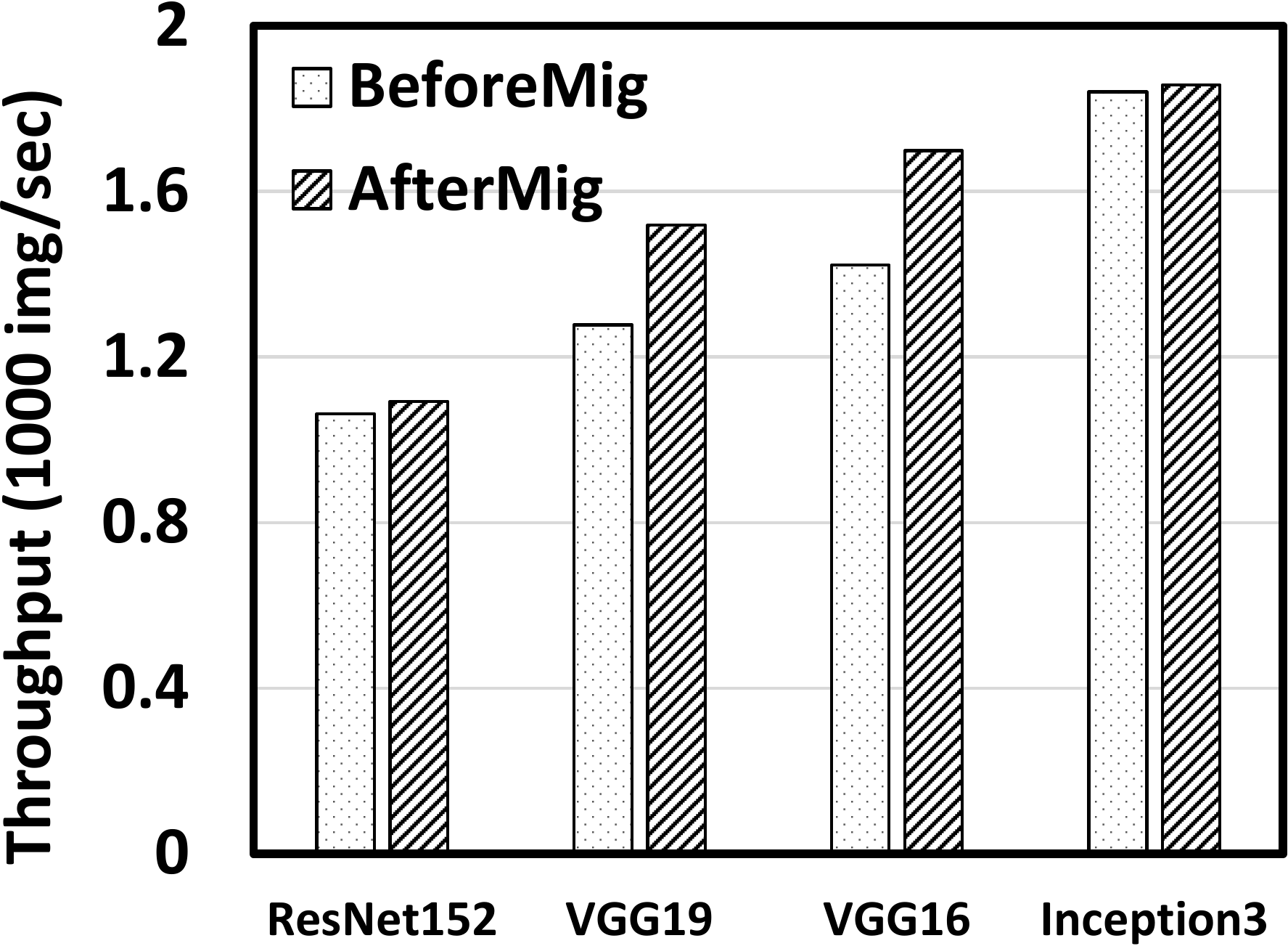}
		\subcaption{Worker migration}\label{fig:migration}
	\end{minipage}
	\begin{minipage}[b]{0.225\textwidth}
		\centering
		\includegraphics[width=\textwidth]{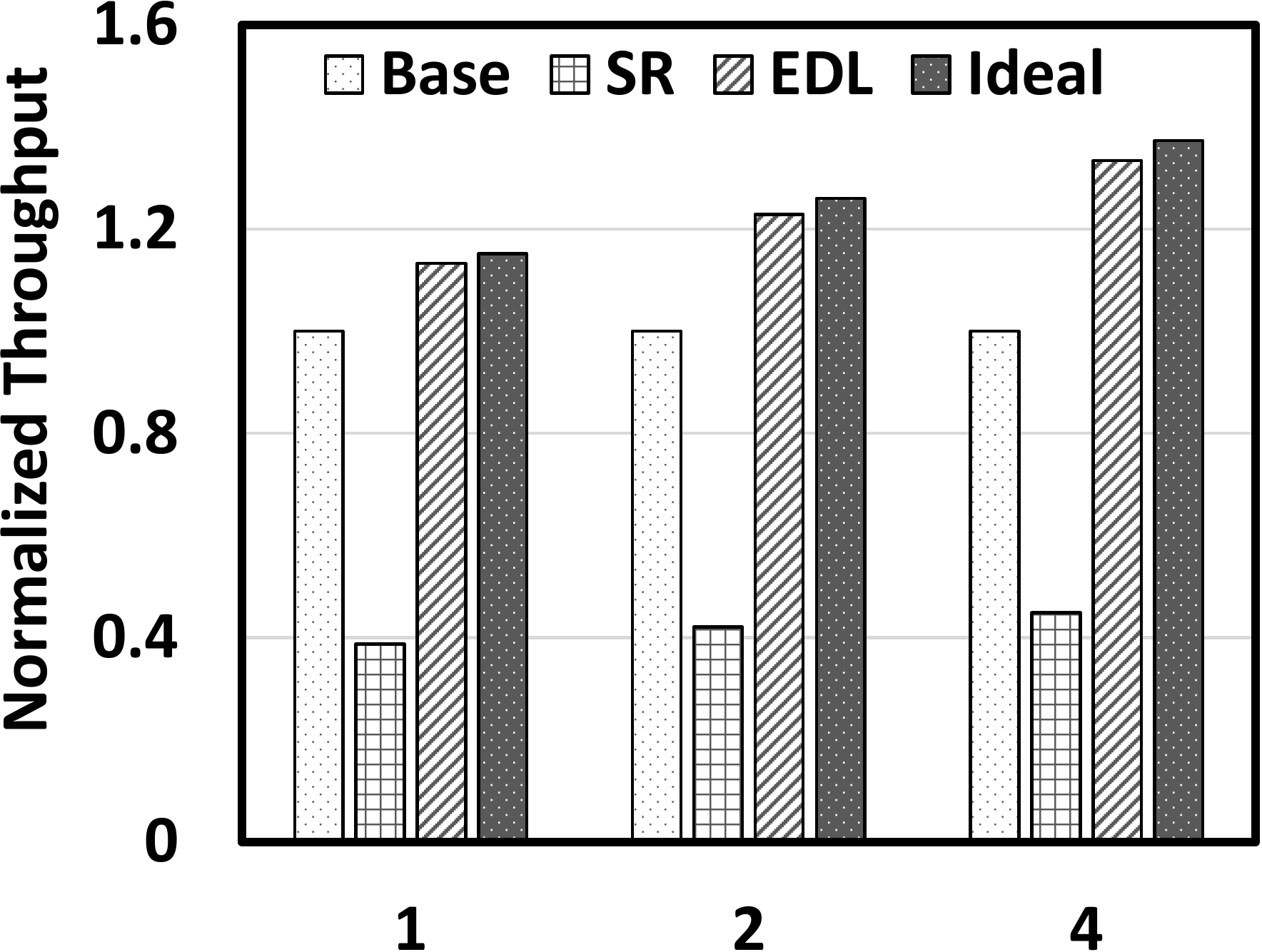}
		\subcaption{Transient GPU usage}\label{fig:transient}
	\end{minipage}
	\vspace{-2mm}	
	\caption{Worker migration and transient GPU usage}
	\label{fig:migration and transient resource}
\end{figure}

\paragraph{Worker migration.} Co-locating GPUs for a job is important for training large models. We considered a job running on 2 machines, each using 4 GPUs. We used EDL to migrate the job to one of the machines and run on the 8 GPUs on that machine. We report the training throughput before and after the migration in Figure~\ref{fig:migration}. For large models, e.g., VGG16 and VGG19, there was an significant increase in throughput (nearly 20\%) after migration, though the increase was not obvious for small models (e.g., 2.9\% for ResNet152). We found that the cost of worker migration was similar to scaling out and training on the target machine is only stopped for less than a second.

\paragraph{Use of transient resources.} To validate the benefit EDL can bring out of the transient idle resources, we conducted an experiment using a job that trained ResNet50 with 4 persistent GPUs and considered the cases that there were 1, 2 and 4 idle GPUs on the same machine. The idle GPUs were revoked every 4 minutes to simulate the transient idle resources reported in~\S\ref{subsec:benifits of elasticity}. Four schemes were used: (1)~\textit{Baseline}, which did not use the idle GPUs and used 4 GPUs for training at all time; 2)~\textit{Stop-resume} (\textit{SR}), which used stop-resume for scaling out and scaling in when using the transient idle GPUs; 3)~\textit{EDL}, which used EDL for scaling; 4)~\textit{Ideal}, which assumed that the scaling completed instantly without any overhead. Note that scaling needed to be conducted twice for each idle interval, i.e., scaling out to add the idle GPUs to training and scaling in to remove these GPUs after the transient period.

Figure~\ref{fig:transient} shows that EDL achieved at least 97\% of the throughput of Ideal. In contrast, stop-resume performed even worse than Baseline due to its high scaling overheads, which is in line with our analysis in~\S\ref{subsec:benifits of elasticity}. We found that 11.7 minutes is the shortest transient interval needed for stop-resume to outperform Baseline with 1 idle GPUs, while EDL only requires the idle interval to be longer than the launch-up time of a worker to outperform Baseline. This result shows that the low scaling overhead enables EDL to utilize idle resources more effectively.

\subsection{Performance on Cluster Scheduling}\label{exp:scheduling}

\begin{figure}[!t]	
	\centering 
	\begin{minipage}[b]{0.23\textwidth}
		\centering
		\includegraphics[width=\textwidth]{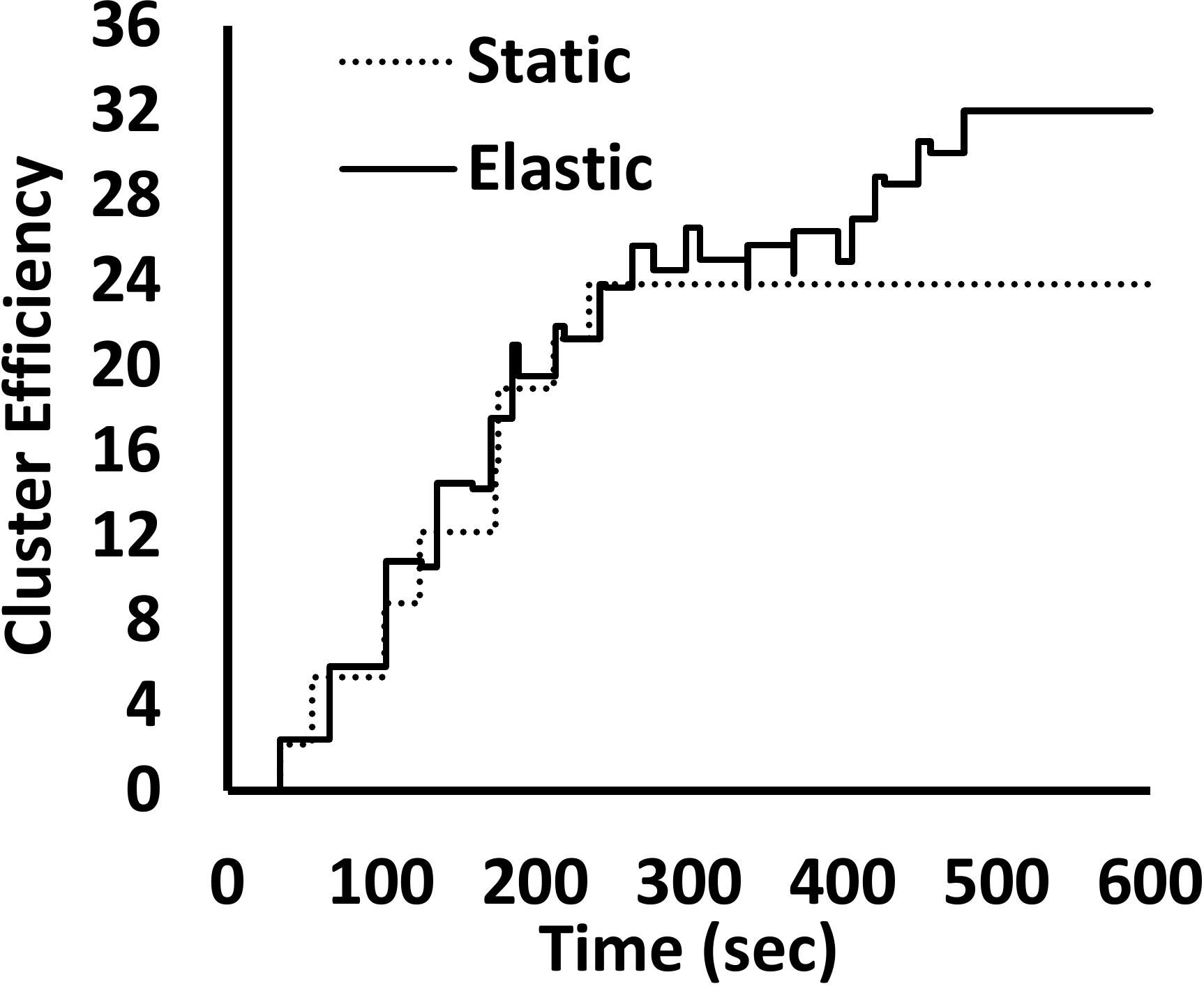}
		\subcaption{Cluster efficency}\label{fig:cluster}
	\end{minipage}
	\begin{minipage}[b]{0.23\textwidth}
		\centering
		\includegraphics[width=\textwidth]{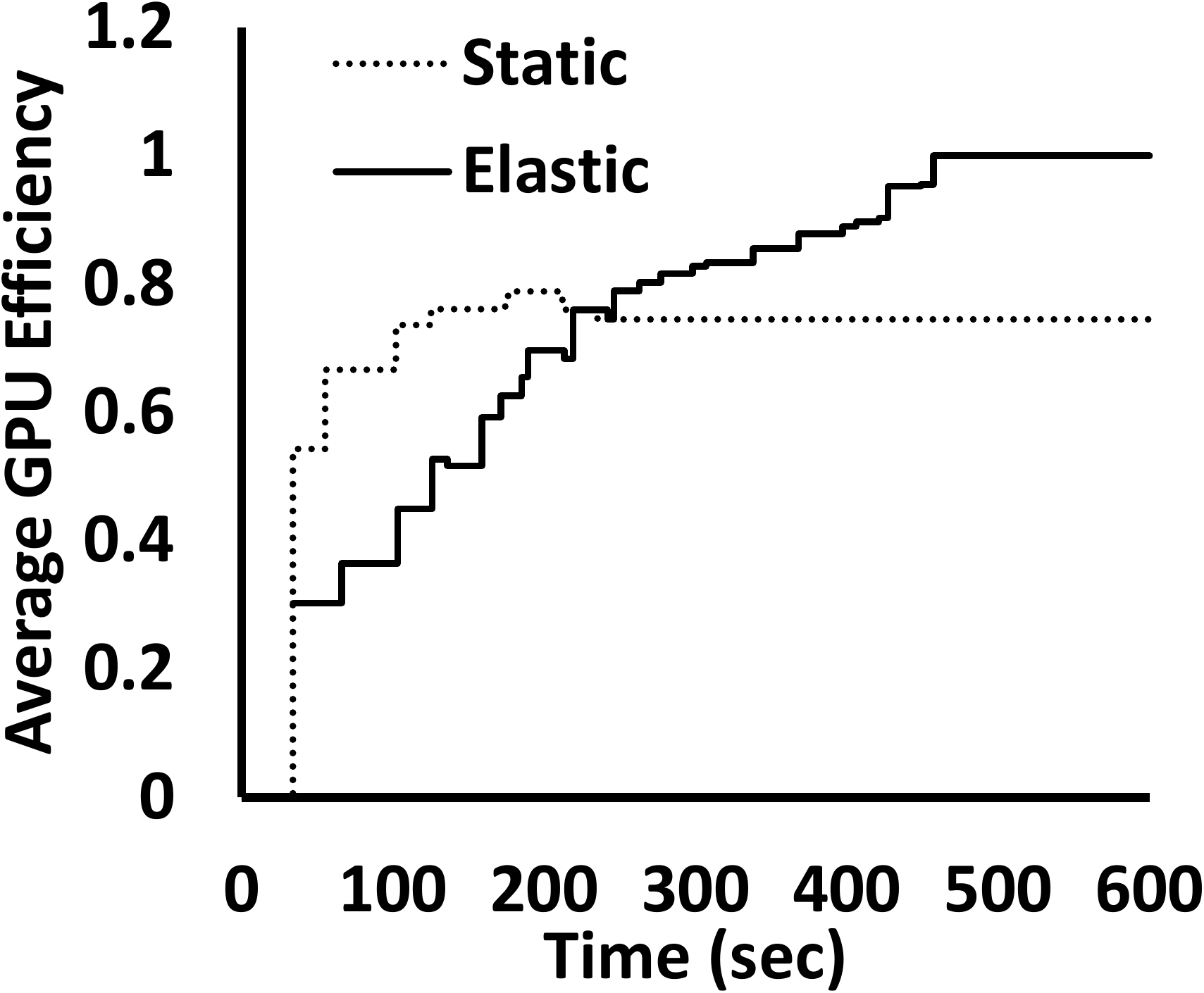}
		\subcaption{Averge GPU efficency}\label{fig:GPU}
	\end{minipage}
	\caption{Performance on synthetic workload}
	\label{fig:synthetic scheduling}
\end{figure}

\paragraph{Synthetic workload.} To demonstrate the benefits of using EDL in scheduling, we created a synthetic workload to evaluate the performance with/without elasticity. We submitted a job to our cluster using 4 machines, each with 8 GPUs, at every 30 seconds, until 16 jobs were submitted  (no job left in the middle). Each job trained a model randomly chosen from the 9 popular DNNs in TensorFlow's official benchmarks \cite{TfCNNBenchmark} (e.g., ResNet, VGG variants) and all jobs ran using 4 GPUs by default. This synthetic workload models different loading conditions that can appear in a production cluster, i.e., the load was low at the beginning when only a few jobs were running, and gradually the cluster was overloaded.

We compared two scheduling strategies: \textit{Static} and \textit{Elastic} (i.e., using EDL). Static ran each job with a static parallelism of 4 and occupied all the GPUs for the first 8 jobs. After that, new jobs were put in a pending queue. Elastic allocated a new job to the least loaded machine (measured by the number of running jobs) and assigned the GPUs on a machine to its jobs uniformly. Elastic also scaled out a job to use any idle GPUs on the machine the job was running, as long as the scale-out does not decrease its throughput~\footnote{We assume profiling was conducted beforehand such that the scheduler knew the performance of the jobs under different parallelism.}. When all GPUs were occupied and a new job was submitted, Elastic scaled in the running job(s) to release GPUs for the new job following the R1 rule introduced in~\S\ref{usecase:scheduling}. 

We report the cluster efficiency and the average GPU efficiency of Static and Elastic in Figure~\ref{fig:synthetic scheduling}. The cluster efficiency is defined as the sum of the per-GPU efficiency for all GPUs, where we set the efficiency of an idle GPU as 0. The average GPU efficiency is the average of the per-GPU efficiency of the active GPUs. Figure~\ref{fig:cluster} shows that Elastic achieved higher cluster efficiency than Static almost all the time, while Figure~\ref{fig:GPU} shows that the per-GPU efficiency of Elastic was lower than Static at the beginning. This is because Elastic scaled out the jobs to use idle GPUs when the cluster load was light, which resulted in lower per-GPU efficiency but higher throughput. The per-GPU efficiency of Elastic became higher than that of Static when half of the jobs were submitted, as it scaled in the jobs to run more jobs concurrently. The small spikes on the curves of Elastic were caused by the scaling operations. Both per-GPU and cluster efficiency of Elastic approached their maximum when 16 jobs were running, while those of Static reached their maximum when approximately 8 jobs were running. The results thus verify that using EDL improves the cluster efficiency under different loading conditions.

\paragraph{Production cluster simulation.}  To show the benefits of using EDL in scheduling a large GPU cluster, we compared Elastic-Tiresias (presented in~\S\ref{usecase:scheduling}) with Tiresias~\cite{GuCSZJQLG19nsdi}. We used the simulator provided in~\cite{GuCSZJQLG19nsdi}, which has been shown to produce results close to actual execution. The simulation was based on the trace data collected from Microsoft's production cluster~\cite{Philly-traces, JeonVPQXY19atc}. The trace data contains more than 100,000 training jobs, but the model architectures of the jobs are not disclosed. Thus, we followed the same approach in~\cite{GuCSZJQLG19nsdi} and generated models chosen uniformly at random from TensorFlow's official benchmarks. Both Tiresias and Elastic-Tiresia were configured with three queues (also called \textit{groups} in~\S\ref{usecase:scheduling}) and the service quantum for $G_0$ and $G_1$ are 500 GPU*sec and 10,000 GPU*sec, respectively. Elastic-Tiresia uses $N\!=\!10$ for the threshold of waiting jobs and $r\!=\!0.5$ for the quality of service guarantee.

\begin{figure}[!t]	
	\centering 
	\begin{minipage}[b]{0.225\textwidth}
		\centering
		\includegraphics[width=\textwidth]{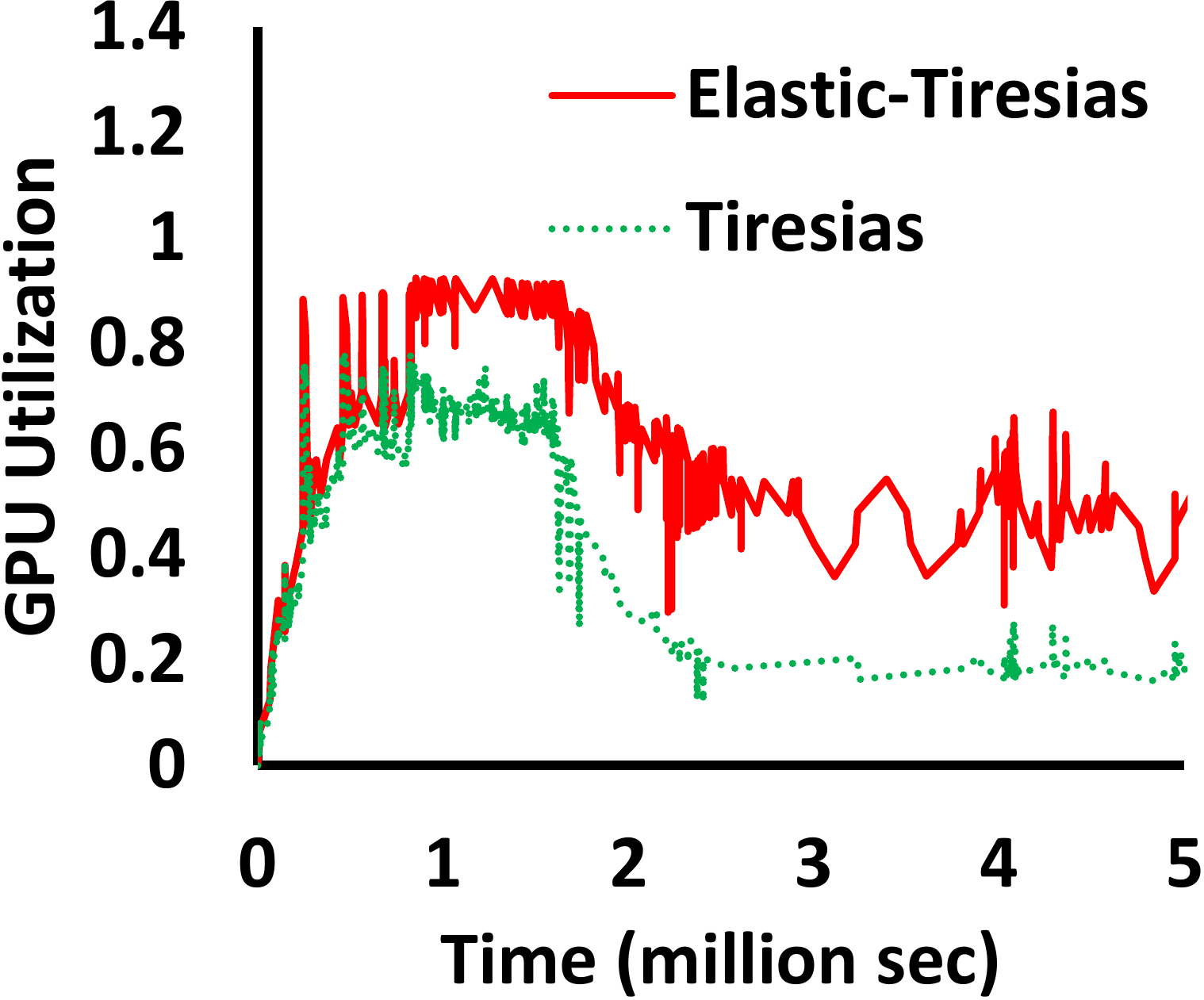}
		\subcaption{GPU ultilization}\label{fig:idle-gpu}
	\end{minipage}
	\begin{minipage}[b]{0.23\textwidth}
		\centering
		\includegraphics[width=\textwidth]{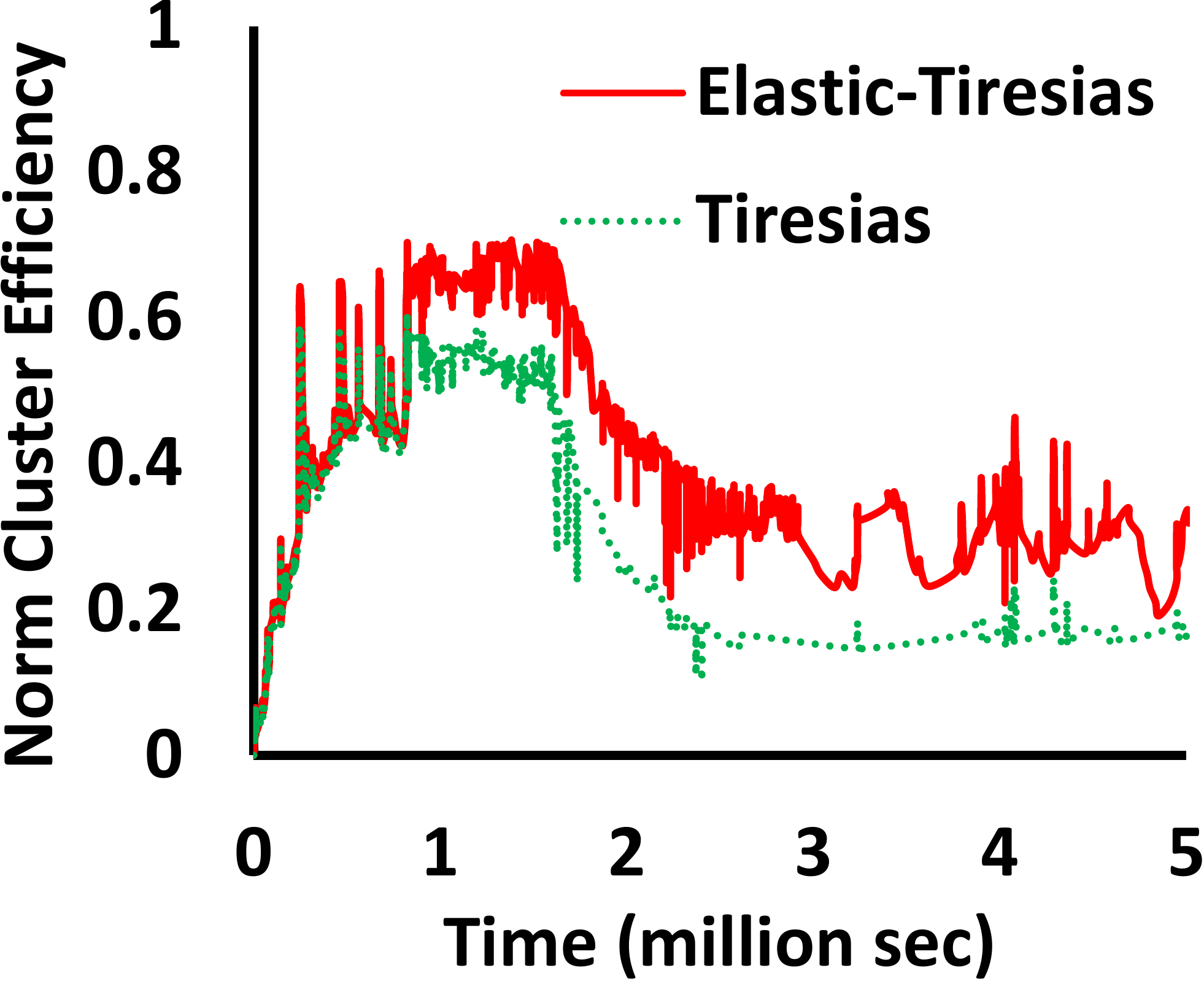}
		\subcaption{Cluster efficency}\label{fig:cluster efficency}
	\end{minipage}
	\caption{Performance on cluster scheduling}
	\label{fig:cluster scheduling}
\end{figure}

We report some statistics of the JCTs of Tiresias and Elastic-Tiresias in Table~\ref{tab:JCT}. With elasticity enabled, the JCTs of Tiresias are significantly reduced. To further examine the scheduling performance of Tiresias and Elastic-Tiresias, we plot the GPU utilization rate (i.e., the fraction of GPUs in use) and the cluster efficiency (normalized by the total number of GPUs) in Figure~\ref{fig:cluster scheduling}. The results show that Elastic-Tiresias achieves higher GPU utilization rate and cluster efficiency than Tiresias. The GPU utilization rate of Elastic-Tiresias is higher because it scales out the jobs to utilize the idle GPUs. The cluster efficiency curve is highly correlated with the curve of the GPU utilization rate, which shows that utilizing the idle GPUs also leads to higher cluster efficiency, which in turn leads to improved JCTs.

\begin{table}[t]
	\centering
	\caption{Statistics of job completion time  (sec)}
	\label{tab:JCT}
	\vspace{-3mm}
	\begin{center}
		\fontsize{8}{9}\selectfont
		\begin{tabular}{ccccc}
			\toprule		
			&  Tiresias        & Elastic-Tiresias & Reduction (\%)\\
			\midrule
			Mean    & 235,068  & 24,658            & 89.5\%    \\
			Median  & 1,080      &   561 	      & 48.1\%    \\
			95th    & 1,914,470 &  88,886        & 95.4\%   \\
			\bottomrule
		\end{tabular}
	\end{center}
\end{table}

\section{Related Work}	\label{sec:related}

\paragraph{Deep learning schedulers.} Instead of using traditional cluster manager such as Yarn\cite{VavilapalliMDAKEGLSSSCORRB13cloud}, Mesos \cite{HindmanKZGJKSS10nsdi}, Omega \cite{SchwarzkopfKAWeurosys13} and Borg\cite{VermaPKOTW15eurosys}, a number of DL-specialized schedulers are proposed for multi-tenant GPU clusters recently, e.g., Optimus~\cite{PengBCWG18eurosys}, Gandiva~\cite{XiaoBRSKHPPZZYZ18osdi} and Tiresias~\cite{GuCSZJQLG19nsdi}. Optimus adjusts the number of parameter servers/workers of MXNet periodically using the stop-resume approach to minimize JCT. Gandiva~\cite{XiaoBRSKHPPZZYZ18osdi} introduces various mechanisms such as~\textit{migration, grow-shrink, profiling and suspend-resume} to adjust resource allocation according to runtime measurements. As~\textit{grow-shrink} adjusts the batch size of a job along with the parallelism, Gandiva only uses it when a job is declared to be parallelism insensitive. As introduced in~\S\ref{usecase:scheduling}, Tiresias approximates the shortest-job-first strategy with a priority discretization framework to alleviate head-of-line blocking. EDL positions itself as a system that provides low-overhead elasticity, and can cooperate with existing GPU schedulers by enabling more frequent parallelism adjustments and supporting scheduling mechanisms such as \textit{migration and profiling} efficiently. EDL also provides consistency semantics under elasticity which helps generalize grow-shrink to all jobs.

\paragraph{Elastic ML/DL systems.} Machine learning (ML) systems are usually based on the parameter-server architecture~\cite{XingHDKWLZXKY15kdd, HuangJWCYYLGC18pvldb, LiAPSAJLSS14osdi, ZhangZXDHLHWXX17atc, XingHDKWLZXKY15kdd} and process distributed ML workloads such as Logistic Regression and Latent Dirichlet Allocation \cite{BleiNJ01nips} in CPU clusters. Elasticity has also been found useful in adapting to resource availability for such workloads. Litz~\cite{QiaoAYCHGX18atc} adopts designs such as update forwarding and executor migration to support the dynamic addition/removal of servers and workers. Based on a performance model, Cruise \cite{lee2019automating} dynamically adjusts the configurations of the parameter servers and workers for optimal performance. 

Baidu's Paddle EDL~\cite{paddle} is a DL system based on the parameter-server architecture and integrated with Kubernetes. Very recently, Ant Financial also introduced an early-stage ElasticDL project~\cite{antfinancial}, which is based on TensorFlow 2.0. Both systems are designed for asynchronous training and fall back to stop-resume if parallelism is adjusted during synchronous training. Concurrent with our work, DL2~\cite{PengBCWMLcorr2019} supports elasticity on (parameter-server-based) MXNet but it is not clear how DL2 hides the overheads of adding new workers and how the training data is partitioned among a dynamic set of workers. Compared with these systems, EDL is based on the Allreduce architecture, supports synchronous training and introduces tailored designs such as stop-free scaling and dynamic data pipeline to reduce the overheads of elasticity.

\paragraph{Systems for transient resources.} Due to the significantly lower price of preemptible instances on cloud than on-demand ones, many systems have been designed to utilize transient resources~\cite{AmazonEC2Spot,GooglePreemptibleInstance}. Proteus~\cite{HarlapTCGG17eurosys} is a parameter server based ML system that manages models on reliable server nodes and allows workers to be dynamically added or removed to utilize the revocable resources. Hourglass~\cite{JoaquimBRM19eurosys} is a graph processing system that partitions a graph into micro partitions and reassigns these micro partitions among the machines when resource changes. Tributary~\cite{HarlapCTGG18atc} runs web servers using transient resources across different cloud markets to avoid  correlated preemptions within one spot market and satisfy quality of service guarantees (e.g., low latency). Flint~\cite{SharmaGHISeurosys16}, Pado~\cite{YangKSLCQCCeurosys17} and TR-Spark~\cite{YanGCGCMcloud16} focus on batch-processing jobs and use smart checkpointing and task scheduling strategies to minimize the impact of resource revocation. While transient workers usually last for hours in cloud spot markets, EDL considers a more stringent situation where it is common that transient GPU resources are only available for minutes, which necessitates elasticity with low overheads.

\section{Conclusions}	\label{sec:end}

We presented EDL, which supports elastic GPU utilization with low overheads. EDL can benefit multi-tenant GPU cluster management in many ways, including improving resource utilization by adapting to load variations, maximizing the use of transient idle GPUs, performance profiling, straggler mitigation, and job migration. We showed in our experiments that significant performance benefits can be obtained using EDL in these applications.

\bibliographystyle{ACM-Reference-Format}
\bibliography{ref}

%
%
%
%
%
%
%
%

\end{document}